%% file: main.tex
\documentclass[letter,11pt]{article}
\pdfoutput=1 

\usepackage[top=1in,bottom=1in,left=0.75in,right=0.75in]{geometry}
\usepackage{amsmath}
\usepackage{amssymb}
\usepackage{amsthm}
\usepackage{mathtools}
\usepackage[only,llbracket,rrbracket]{stmaryrd}
\usepackage[scr=boondoxo]{mathalfa} 
\usepackage[english]{babel}
\usepackage{graphicx}
\usepackage[font=small,figurewithin=section]{caption}
\usepackage{subcaption}
\usepackage{float}
\usepackage{hyperref}
\usepackage[all]{xy}
\usepackage{multirow}
\usepackage{array}
\usepackage{enumitem}
\usepackage{tikz}
\usetikzlibrary{arrows.meta}
\usepackage{tikz-3dplot}
\usetikzlibrary{matrix,arrows,decorations.pathmorphing}

\usepackage{pgfplots}
\pgfplotsset{compat=newest}

\usepackage{color}
\usepackage{authblk}
\usepackage[mathscr]{euscript}

\usepackage{relsize}

\usepackage{comment}

\numberwithin{equation}{section}
\makeatletter
\g@addto@macro\bfseries{\boldmath}
\makeatother

\hypersetup{colorlinks=true,citecolor=black,filecolor=black,linkcolor=black,urlcolor=black}
\hypersetup{pdfstartview=FitB,pdfpagemode=UseNone}

\setlist{nolistsep}

\newcolumntype{L}[1]{>{\raggedright\let\newline\\\arraybackslash\hspace{0pt}}m{#1}}
\newcolumntype{C}[1]{>{\centering\let\newline\\\arraybackslash\hspace{0pt}}m{#1}}
\newcolumntype{R}[1]{>{\raggedleft\let\newline\\\arraybackslash\hspace{0pt}}m{#1}}
\newcolumntype{N}{@{}m{0pt}@{}}
\SetSymbolFont{stmry}{bold}{U}{stmry}{m}{n}
\usepackage{derivative}
\usepackage{algorithm}
\usepackage{algpseudocode}
\usepackage{algorithmicx}
\algrenewcommand\algorithmicrequire{\textbf{Input:}}
\algrenewcommand\algorithmicensure{\textbf{Output:}}

\usepackage[backend=bibtex,
	giveninits=true,
	maxbibnames=99,
	date=year,
	doi=true,
	url=false,
	isbn=false,
	eprint=false]{biblatex}
\usepackage{csquotes}

\newcommand{\revisedA}[1]{\textcolor{black}{#1}}
\newcommand{\revisedB}[1]{\textcolor{black}{#1}}

\newcommand{\revA}{\revisedA}
\newcommand{\revB}{\revisedB}

\newcommand{\revS}[1]{\textcolor{black}{#1}}

\input{preamble}

\input{my_preamble}
\begin{document}
%
\title{Bent optical waveguide finite element analysis\\ with a 3D envelope Maxwell model}


\author[1,2]{Jaime Mora-Paz\thanks{e-mail: jdmorap@utexas.edu, jaimed.morap@konradlorenz.edu.co}}
\author[1]{Stefan Henneking}
\author[1]{Leszek Demkowicz}
\author[3]{Jacob Grosek}
\affil[1]{\scriptsize Oden Institute for Computational Engineering and Sciences, The University of Texas at Austin, 201 E 24th St, Austin, TX 78712, USA}
\affil[2]{\scriptsize Facultad de Matemáticas e Ingenierías, Fundación Universitaria Konrad Lorenz, Cr 9 Bis No. 62-43, Bogotá, Colombia}
\affil[3]{\scriptsize Information \& Spectrum Warfare Directorate, Air Force Research Laboratory, 3550 Aberdeen Ave SE, Albuquerque, NM 87117, USA}
\date{\vspace{-10mm}}
\maketitle

\begin{abstract}
With the goal of accurately extracting the optical field losses in a three-dimensional (3D), circularly coiled waveguide (e.g., bent optical fiber), this effort presents the numerical methodologies that are implemented for an envelope Maxwell model that propagates electromagnetic fields as an entirely boundary value problem. 
Our unique modeling approach includes an ultraweak variational formulation of the envelope Maxwell model in the curved geometry of the bending, which is discretized by the discontinuous Petrov–Galerkin (DPG) method, which permits residual-driven mesh and polynomial-order adaptivity. 
This also, then, requires a unique approach for constructing perfectly matched layers (PMLs) as absorbing boundary conditions in both the direction of optical field propagation and in the tangential directions, where unguided energy escapes the waveguide. 
Our coiled waveguide modeling technology extracts the mode confinement losses from the propagation of the coherent optical field through the bent waveguide. 
We verify our simulations against the semi-analytical results from the analogous bent slab waveguide problem, and we successfully demonstrate stable convergence to loss values for the 3D coiled optical fiber problem, which has never been done previously for our specific modeling approach. 
\end{abstract}



\subsection*{Acknowledgments}  L. Demkowicz, J. Mora-Paz and S. Henneking  were supported by AFOSR grant FA9550-23-1-0103. The authors acknowledge the Texas Advanced Computing Center (TACC) for providing computational resources that have contributed to the research results reported within this paper.


\input{1_intro}

\input{2_model}
\input{3_discretization}

\input{4_experiments}

\input{5_conclusions}

\section*{Disclaimers}
This article has been approved for public release; distribution unlimited. 
Public Affairs release approval {\#}AFRL-2026-1401. 
The views expressed in this article are those of the authors and do not necessarily reflect the official policy or position of the Department of the Air Force, of the Department of Defense, nor of the U.S. government.

\printbibliography[heading=bibintoc]
\setcounter{section}{0}

\end{document}

%% file: preamble.tex
\definecolor{ICES}{RGB}{94, 156, 174}

\definecolor{ORANGE}{RGB}{191, 87, 0}

\definecolor{RED}{RGB}{190, 30, 49}

\definecolor{SUN}{RGB}{227, 81, 51}

\definecolor{GREEN}{RGB}{0, 171, 86}

\definecolor{BLUE}{RGB}{11, 78, 179}

\definecolor{BROWN}{RGB}{122, 80, 40}

\definecolor{GREY}{RGB}{50, 50, 50}

\definecolor{TEAL}{RGB}{0, 160, 176}

\definecolor{AFBlue}{RGB}{13, 77, 140}

\definecolor{AFDarkGray}{RGB}{89, 90, 89}

\definecolor{AFLightBlue}{RGB}{0, 188, 228}

\definecolor{AFLightGray}{RGB}{203, 204, 203}

\definecolor{AFRed}{RGB}{179, 40, 45}

\definecolor{AFYellow}{RGB}{251, 206, 32}

\definecolor{darkplum}{RGB}{78, 26, 104}

\definecolor{dgreen}{rgb}{0.2, 0.5, 0.2}

\definecolor{dorange}{rgb}{1, 0.55, 0}

\definecolor{dorchid}{rgb}{0.6, 0.196, 0.8}

\definecolor{gold}{rgb}{0.8, 0.498, 0.196}

\definecolor{ultramarineblue}{RGB}{65, 102, 245}

\definecolor{deepskyblue}{RGB}{0, 191, 255}    

\definecolor{blueviolet}{RGB}{138, 43, 226}

\newcommand{\rbend}{r_{0}}
\newcommand{\rcore}{\rho_{\text{core}}}
\newcommand{\rclad}{\rho_{\text{clad}}}
\newcommand{\rfi}{\rho_{\text{fiber}}}

\newcommand{\ncore}{n_{\text{core}}}
\newcommand{\nclad}{n_{\text{clad}}}
\newcommand{\ncoat}{n_{\text{coat}}}

\newcommand{\nair}{n_{\text{air}}}

\newcommand{\pOne}{p_{11}^{\text{glass}}}
\newcommand{\pTwo}{p_{12}^{\text{glass}}}
\newcommand{\PR}{\nu_{\text{glass}}}
\newcommand{\ntensor}{\boldsymbol{\sfn}}
\newcommand{\coil}{\text{coil}}
\newcommand{\straight}{\text{str}}
\newcommand{\dncoil}{\delta\ntensor_{\coil}}

\newcommand{\defeq}{\coloneqq}

\newcommand{\R}{\mathbb{R}} 
\newcommand{\Z}{\mathbb{Z}} 
\newcommand{\C}{\mathbb{C}} 
\newcommand{\zI}{\mathfrak{i}} 

\newcommand{\idmatrix}{\textup{\uppercase\expandafter{\romannumeral 1}}}
\newcommand{\mesh}{\mathcal{T}} 
\newcommand{\T}{\mathsf{T}} 
\newcommand{\bdry}{\partial} 
\newcommand{\tr}{\mathrm{tr}} 
\newcommand{\diag}{\mathrm{diag}} 

\newcommand{\bveps}{\boldsymbol{\varepsilon}}

\DeclareMathOperator{\curl}{curl}
\let\div\relax 
\DeclareMathOperator{\div}{div}

\newcommand{\idtens}{\mathbf{I}} 

\newtheoremstyle{boldremark}
    {\dimexpr\topsep/2\relax} 
    {\dimexpr\topsep/2\relax} 
    {}          
    {}          
    {\bfseries} 
    {.}         
    {.5em}      
    {}          
\theoremstyle{definition}
\newtheorem*{definition*}{Definition}
\theoremstyle{plain}

\theoremstyle{boldremark}
\newtheorem{remark}{Remark}
\theoremstyle{definition}

\makeatletter
\newcommand{\dashto}[1][2pt]{
  \settowidth{\@tempdima}{${}\rightarrow{}$}
  \makebox[\@tempdima]{${}\rightarrow{}$}
  \makebox[-\@tempdima]{\hspace{-0.1\@tempdima}\color{white}\rule[0.5ex]{#1}{1pt}}
  \makebox[\@tempdima]{}
  }
\makeatother
\let\tilde\widetilde

\newcommand{\mcE}{\mathcal{E}}

\newcommand{\mcH}{\mathcal{H}}

\newcommand{\mcO}{\mathcal{O}}
\newcommand{\mcP}{\mathcal{P}}

\newcommand{\sfk}{\mathsf{k}}

\newcommand{\sfn}{\mathsf{n}}

\newcommand{\sft}{\mathsf{t}}
\newcommand{\sfu}{\mathsf{u}}
\newcommand{\sfv}{\mathsf{v}}

\newcommand{\sfA}{\mathsf{A}}

\newcommand{\sfE}{\mathsf{E}}
\newcommand{\sfF}{\mathsf{F}}
\newcommand{\sfG}{\mathsf{G}}
\newcommand{\sfH}{\mathsf{H}}

\newcommand{\sfK}{\mathsf{K}}

\newcommand{\sfT}{\mathsf{T}}

\newcommand{\scU}{\mathscr{U}}
\newcommand{\scV}{\mathscr{V}}

\newcommand{\nml}{\mathbf{n}}

\newcommand{\bfe}{\mathbf{e}}

\usepackage{mathtools}

\makeatletter
\DeclareRobustCommand\widecheck[1]{{\mathpalette\@widecheck{#1}}}
\def\@widecheck#1#2{%
    \setbox\z@\hbox{\m@th$#1#2$}%
    \setbox\tw@\hbox{\m@th$#1%
       \widehat{%
          \vrule\@width\z@\@height\ht\z@
          \vrule\@height\z@\@width\wd\z@}$}%
    \dp\tw@-\ht\z@
    \@tempdima\ht\z@ \advance\@tempdima2\ht\tw@ \divide\@tempdima\thr@@
    \setbox\tw@\hbox{%
       \raise\@tempdima\hbox{\scalebox{1}[-1]{\lower\@tempdima\box
\tw@}}}%
    {\ooalign{\box\tw@ \cr \box\z@}}}
\makeatother

\newcommand{\bff}{\mathbf{f}}
\newcommand{\bfd}{\mathbf{d}}
\newcommand{\bfl}{\mathbf{l}}
\newcommand{\bfn}{\mathbf{n}}

\newcommand{\bfr}{\mathbf{r}}

\newcommand{\bfB}{\mathbf{B}}

\newcommand{\bfG}{\mathbf{G}}

\newcommand{\bsfK}{\boldsymbol{\sfK}}

\newcommand{\bzero}{\mathbf{0}}

\newcommand{\lift}{{\textrm{lift}}}
\newcommand{\bc}{{\textrm{b.c.}}}
\newcommand{\PML}{{\textrm{PML}}}

\newcommand{\pmlparameter}{C_{\PML}}

\newcommand{\bsveps}{\boldsymbol{\varepsilon}}

\renewcommand{\Re}{\operatorname*{Re}}
\renewcommand{\Im}{\operatorname*{Im}}

\newcommand{\bsmu}{\boldsymbol{\mu}}

\usepackage{tikz}
\usetikzlibrary{shapes.geometric, shapes.misc, arrows}
\tikzstyle{startstop} = [rectangle, rounded corners, minimum width=1.8cm, minimum height=0.5cm,text centered, draw=black, fill=red!30]
\tikzstyle{io} = [trapezium, trapezium left angle=70, trapezium right angle=110, minimum width=1.8cm, minimum height=0.5cm, text centered, draw=black, fill=blue!30]
\tikzstyle{process} = [rectangle, minimum width=1.8cm, minimum height=0.5cm, text centered, draw=black, fill=orange!30]
\tikzstyle{decision} = [diamond, minimum width=1.8cm, minimum height=0.5cm, text centered, draw=black, fill=green!30]
\tikzstyle{forloop} = [chamfered rectangle, chamfered rectangle xsep = 0.5cm, text centered, draw=black, fill=teal!30]
\tikzstyle{cont} = [circle, minimum width=0.3cm,draw=black,fill=teal!30]
\tikzstyle{arrow} = [thick,->,>=stealth]

\usepackage{listings}
\usepackage{lscape}
\usepackage{xfrac}
\usepackage{xcolor}
\usepackage{cancel}

%% file: my_preamble.tex
\def\be{\begin{equation}}
\def\ee{\end{equation}}
\def\ba{\begin{array}}
\def\ea{\end{array}}
\def\bea{\begin{eqnarray}}
\def\eea{\end{eqnarray}}
\def\beas{\begin{eqnarray*}}
\def\eeas{\end{eqnarray*}}

%% file: 1_intro.tex
\section{Introduction}
\label{sec:intro}

\begin{figure}[htb]
    \centering
    \includegraphics[angle=0,width=0.5\textwidth]{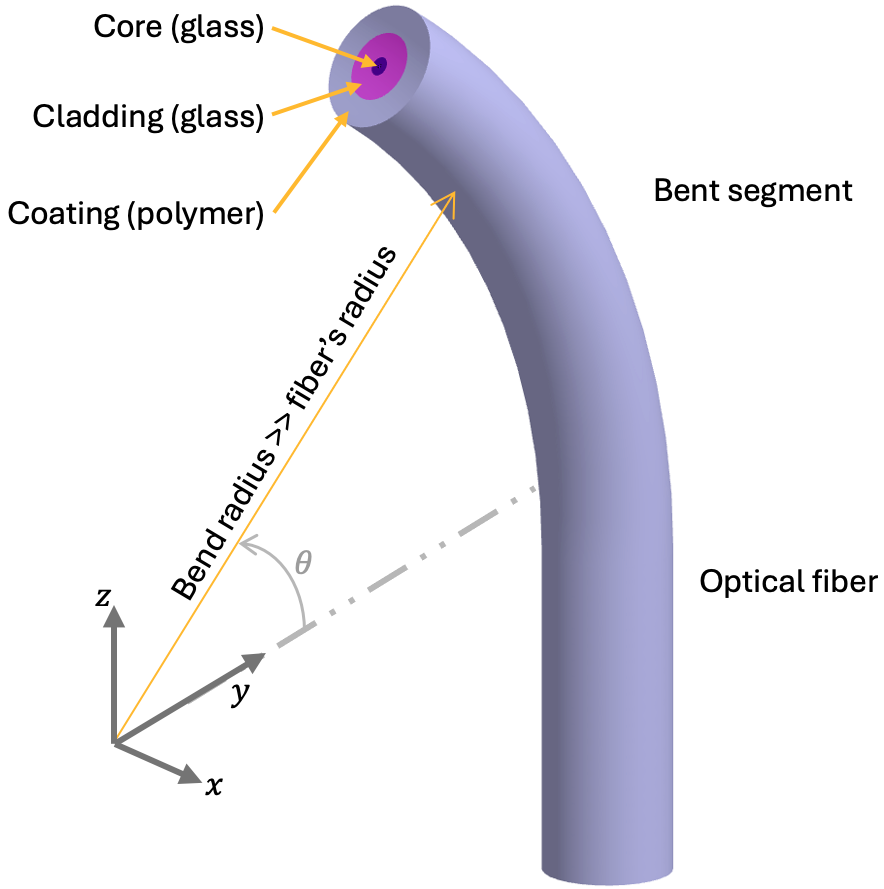}\\
    \caption{This diagram depicts a straight optical fiber (lower segment) fused (spliced) seamlessly with a circularly bent optical fiber (upper segment). This allows one to launch the optical field into the perfectly guiding (i.e.\ lossless) straight segment so that it can propagate into the coiled segment, where it is no longer perfectly guiding and thus experiences confinement loss. For realistic portrayals of bent waveguides, the fiber's radius is much smaller than the bend radius: $\rfi \ll \rbend$.}
    \label{fig:basic-diagram-bent-fiber}
\end{figure}

\paragraph{Description.}
The present work is dedicated to the development and numerical demonstration of high-fidelity finite element discretizations for wave propagation problems in a circularly coiled (or bent) optical waveguide. 
Figure~\ref{fig:basic-diagram-bent-fiber} illustrates the scenario of interest, where an optical fiber waveguide, of cylindrical shape, transitions into a toroidal body when subjected to bending. 
The center of the coordinate system is set at the center of the circular coiling, the bending occurs in the $yz$-plane, and the polar angle, $\theta$, is measured in the counter-clockwise direction, starting on the $y$-axis. 
The \emph{splice} point between the straight portion of the fiber and the bent segment is placed at $\theta = 0$, and the \emph{bend radius}, $\rbend$, measured from the origin to the center of the bent waveguide segment, is much greater than the fiber's radius, $\rfi$ (e.g., while $\rfi$ is a fraction of a millimeter, $\rbend$ is typically on the order of centimeters or tens of centimeters).

\begin{figure}[htb]
    \begin{subfigure}{0.44\textwidth}
        \centering
        \footnotesize
        \begin{tikzpicture}[ scale=0.72, every node/.style={scale=0.64}]
            \def\rcoreval{0.55}
            \def\rcladval{2.0}
            \def\rcoatval{4.0}
            \draw[fill=cyan!10] (0,0) circle(\rcoatval);
            \draw[fill=cyan!30] (0,0) circle(\rcladval);
            \draw[fill=cyan!50] (0,0) circle(\rcoreval);
            \draw[->,black] (-\rcoatval - 0.5, 0) -- (\rcoatval + 0.5, 0) node[anchor = west] {};
            \draw[->,black] (0, -\rcoatval - 0.5) -- (0, \rcoatval + 0.5) node[anchor = south] {};
            \draw[->] (0,0) -- ({\rcoreval * cos(130)}, {\rcoreval * sin(130)}) node[pos = 1.2,above] {\Large $\rcore$};
            \draw[->] (0,0) -- ({\rcladval * cos(5)}, {\rcladval * sin(5)}) node[pos = 0.75,above] {\Large $\rclad$};
            \draw[->] (0,0) -- ({\rcoatval * cos(-20)}, {\rcoatval * sin(-20)}) node[pos = 0.9,above] {\Large $\rfi$};

            \draw[->, blue] (0,0) -- ({1.35 * \rcladval * cos(50)}, {1.35 * \rcladval * sin(50)}) node[anchor = west] {\large $\rho$};
            \draw[->, blue] (0,1.1 * \rcladval) arc[start angle=90, end angle = 50, radius = 1.1 * \rcladval];
            \node[blue] at ({1.2 * \rcladval * cos(70)}, {1.2 * \rcladval * sin(70)}) {\large $\varphi$};
            
            \node[magenta] at (0.0,-0.3) {\large core};
            \node[magenta] at (-0.8,-1.0) {\large cladding};
            \node[magenta] at (-0.7,-3.0) {\large coating};
        \end{tikzpicture}
        \label{subfig:StraightFiberCrossSection}
    \end{subfigure}
    %
    \begin{subfigure}{0.55\textwidth}
        \includegraphics[angle=0,width=\linewidth]{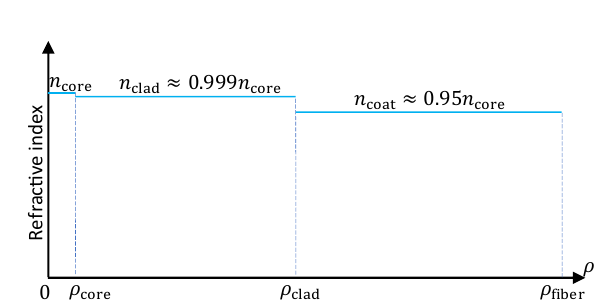}
        \label{subfig:StraightFiberRefractiveIndex}
    \end{subfigure}
    \caption{Cross-section of a straight, step-index optical fiber waveguide: (left) schematic of geometrical sizes and (right) its corresponding radially dependent refractive index profile. }
    \label{fig:cross-section-schematic}
\end{figure}
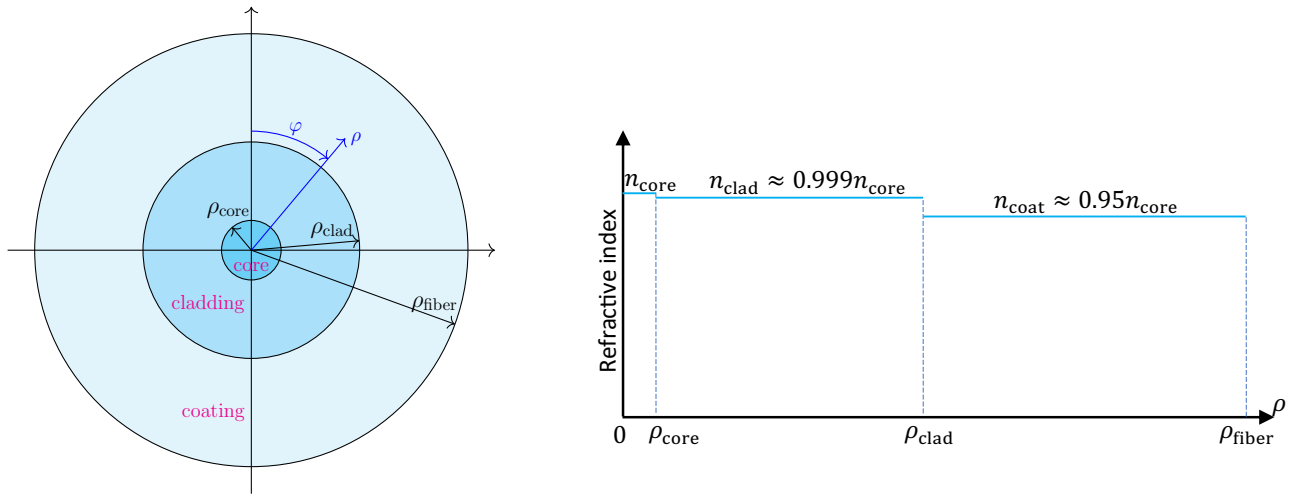

The optical fiber is typically \revB{composed of} three radially separated layers: the core region (innermost material, which is where the majority of the optical field is guided), the cladding (intermediate layer), and the coating (outermost layer). 
The latter is made of polymer, while the inner two layers are \revB{made of} fused silica glass. 
Figure~\ref{fig:cross-section-schematic} displays the cross-sectional details of the fiber. 
The left-side drawing depicts the local transverse coordinates, $(\varphi, \rho)$, along with the characteristic dimensions of the fiber: the core radius, $\rcore$, usually on the order of tens of microns; the cladding radius, $\rclad \approx 10\rcore$, and the fiber/coating radius, $\rfi \approx \rclad + (50\text{--}100 \ \mu$m). 
The right-side schematic of Figure~\ref{fig:cross-section-schematic} portrays the approximate radial distribution of the refractive index value.
This common double-clad \textit{step-index} design has its highest refractive index as $\ncore \sim 1.45$ in the core region, followed by a slightly lower index in the cladding region, $\nclad$ (with a difference \revB{on} the order of $\ncore - \nclad \sim 10^{-3}$). 
Finally, the polymer coating that surrounds the glass cladding has a smaller refractive index, $\ncoat \approx 0.95 \nclad$. 
In a straight fiber, the optical field is highly concentrated within the core region and rapidly attenuates radially. 
Within the coating layer, the propagating optical field is practically negligible. 
However, when the fiber is coiled, the transmitted light may, at least partially, escape the core region, possibly finding its way to the edge of, if not into, the polymer jacket (depending on the optical field's wavelength, the bend radius, and the fiber design). 

In particular, we want to be able to capture the variation of the electromagnetic field as it moves along the bent waveguide, along with the radiation losses that are known to occur due to the imposed curvature. 
Thus, we need a numerical approach that is able to reliably handle the three-dimensional physics \revB{involved and} robustly solve the problem in a sufficiently long domain so as to clearly perceive the effects of the curvature, which implies resolving wave propagation over a very large number of wavelengths. 
For a finite element discretization, this means that local refinements are needed, especially outside of the core region. 
Additionally, the numerical formulation must accurately represent the radiation condition and the truncation of the waveguide in the direction of propagation.

\revB{The proposed methodology is based on (1)~the adaptation of an envelope Maxwell model for coiled waveguides, which is well-suited for simulating high-frequency wave propagation in long domains; (2)~the derivation of an ultraweak variational formulation for the boundary value problem (BVP), which reduces the dispersive effects of numerical pollution in wave propagation problems; (3)~the discretization of the BVP with the discontinuous Petrov--Galerkin (DPG) method, which provides stability and an error indicator useful for automated mesh adaptivity; and (4)~the implementation within a specialized $hp$-finite-element software suite on high-performance computing resources to obtain high-fidelity results.}

\paragraph{Background.}
This paper is a continuation of previous efforts to derive, implement, and analyze Maxwell formulations for high-fidelity optical fiber modeling.
Previous work developed DPG formulations of 3D Maxwell optical fiber models that simulate passive Raman gain amplification~\cite{nagaraj2018raman} and active gain amplification via ytterbium-doped fiber cores~\cite{henneking2021fiber, henneking2025envelope}; implemented these models for large-scale parallel computing~\cite{henneking2021phd, henneking2022parallel} in the $hp$3D finite element library~\cite{henneking2024hp3d, mora2019fast, badger2020fast}; studied the thermally-induced transverse mode instability nonlinearity~\cite{henneking2026comparison}; and developed an accompanying stability analysis for acoustic and electromagnetic waveguides~\cite{henneking2021pollution, melenk2025waveguide1, demkowicz2024waveguide2, demkowicz2024waveguide3}.
The 3D models using our modeling methodology \revB{have been developed and implemented exclusively} for \emph{straight} (i.e., unbent) fiber configurations.
Building on these previous efforts, this paper extends the existing Maxwell models---in particular, the envelope Maxwell formulation described in \cite{henneking2025envelope}---from the straight waveguide setting to the bent waveguide.
This extension requires (1)~modification of the vectorial envelope ansatz to recognize the transformation of the wave propagation direction from the straight (longitudinal / $z$) Cartesian coordinate to the bent (arc / $\theta$) toroidal coordinate; and (2)~construction of a \emph{perfectly matched layer} (PML) to absorb outgoing waves in the radial (transverse) directions.
Finally, although not explicitly addressed in the present work, in addition to geometric effects from coiling, the propagating electromagnetic field in curved waveguides is also impacted by elasto-optical effects (i.e.\ anisotropy in the refractive index caused by mechanical stresses) induced by bending, torsion, and tension \cite{zubia1997theoretical,tai2002optical}.

We are also building on our previous work, which analyzed the mode confinement losses of circular slab waveguides using a semi-analytical method \cite{mora2025bessel}. 
In this effort, we verify the reliability of our proposed approach by \revB{comparing the mode loss obtained from field propagation through the coiled waveguide with} the loss value directly extracted from the eigensolution for the mode data. 
These semi-analytic solutions exist for the 2D bent slab waveguide, but not for the 3D coiled optical fiber waveguide.

One important application for waveguide bending is to \revB{filter out} unwanted higher-order modes (HOMs) from large-mode-area (LMA; i.e.\ \revB{multi-mode}) fibers~\cite[{\S}5C]{richardson2010high}. 
Indeed, the best output beam quality for an optical fiber occurs when the transmitted optical field energy is entirely contained within the waveguide's fundamental mode (FM)~\cite{yoda2006beam}. 
For example, it is known that \textit{differential bend loss}, which is a reference to the fact that HOMs tend to experience much larger confinement losses than \revB{the corresponding FM does} for a fixed bend radius ($\rbend$), helps suppress the onset of the thermally-induced transverse mode instability (TMI) nonlinearity~\cite[{\S}6.1]{jauregui2020tmi}. 
For most applications, TMI is seen as a detrimental problem to be avoided. 
Additionally, coiling a waveguide can be beneficial for compact packaging purposes~\cite{taylor2003bending} and as a means of effective cooling by wrapping the waveguide onto a spool/mandrel, which acts as a heat sink~\cite[{\S}2.2.3.1]{power2009mccomb}.

\paragraph{Research goals and scope.}

Although the analysis of a circularly coiled optical fiber problem has been well studied in previous efforts~\cite{marcuse1976curvature, marcuse1976field, marcuse1982influence, hobbs1990loss, deng1998investigation, feng2002computation, smink2007bending, schermer2007improved, eguchi2010bending, xiao2012vector, wu2021full}, our effort is unique because we are discretizing the entire fiber domain with 3D finite elements such that the optical field propagation is captured by a full BVP framework.

Furthermore, the \textit{envelope} formulation of the Maxwell problem, posed as an ultraweak variational formulation that is discretized with the DPG method, is a relatively new concept that is integral to reducing the computational costs sufficiently so as to allow for longer propagation distances of the high-frequency wave (in this case, propagating over thousands of wavelengths). 
The DPG method allows for residual-driven mesh refinements, which will be shown to be essential for obtaining an accurate solution to the problem. 
The numerical techniques for \revB{setting up} this coiled 3D waveguide problem, including \revB{the treatment of} PML boundary conditions, require thoughtful implementation and are novel contributions to the optical fiber simulation field. 
Our goal is then to delineate the formulation of the bent optical fiber for our modeling approach and to demonstrate its success in capturing the confinement loss of the launched optical field within the waveguide.

Therefore, the contributions of this article include: 
\begin{enumerate}
    \item Extension of \revB{the} Maxwell envelope formulation to the bent waveguide/fiber setting.
    \item PML construction for circularly coiled waveguides.
    \item Implementation of an ultraweak DPG discretization for the bent waveguide's envelope Maxwell model.
    \item Numerical verification of power losses due to geometric bending effects in a slab waveguide.
    \item First example of an adaptive finite element method for the 3D Maxwell simulation of bent optical fibers.
\end{enumerate}
\revB{The paper is organized in the following way: Section~\ref{sec:model} presents the model; Section~\ref{sec:discretization} describes the discretization and implementation; Section~\ref{sec:experiments} presents the numerical experiments; and Section~\ref{sec:conclusions} discusses the conclusions and future work.}

%% file: 2_model.tex
\section{Proposed model}
\label{sec:model} 
The \revB{proposed} optical fiber vectorial Maxwell model is based on the \emph{full envelope ansatz} methodology. 
We strongly encourage the readers to first familiarize themselves with our earlier work on this subject \cite{henneking2025envelope}, which was presented for the straight waveguide setting. 
In this effort, the model formulation considers the curvilinear geometry of the coiled waveguide. 
Once the governing vector field equations are established, we determine the treatment of the boundary conditions (via PMLs) \revB{and} finally arrive at the strong and \emph{ultraweak} forms of our \revB{BVP}.

\subsection{Envelope Maxwell model for a circularly bent waveguide}

Let us start with Maxwell's equations in their time-harmonic form\revisedB{, defined over a domain $\Omega\subset\R^3$}:
\begin{equation}
    \left\{
\begin{array}{rll}
    \nabla\times E 
    + \zI \omega \bsmu H & = \bzero 
    & \text{ (Faraday's law),} \\
    \nabla \times H 
    - \zI \omega \bsveps E & = \bzero 
    & \text{ (Amp\`{e}re--Maxwell's law),}
\end{array}
\right.
\label{eq:time-harmonic-Maxwell}
\end{equation}
where \revisedB{the unknowns are the vector fields $E$ and $H$, which stand for the electric and magnetic fields, respectively. We} have implicitly employed the time-dependence ansatz factor $e^{+\zI\omega t}$\revisedB{, where $\zI$ is the imaginary unit, $\omega$ is the angular frequency, and $t$ is the time.}  Here, we regard the electric permittivity, $\bsveps$, and the magnetic permeability, $\bsmu$, as complex-valued tensors rather than scalars, which may depend on \revisedB{$\omega$ and on the medium (via the position $(x,y,z)\in\Omega$}).
Throughout this effort, the optical field is presumed to be monochromatic (single-frequency\revisedB{, constant $\omega$}); e.g., a laser field. 
Therefore, the time-harmonic ansatz removes all time dependence from our problem. \revisedB{Note that in \eqref{eq:time-harmonic-Maxwell} we only consider homogeneous loads in both equations of the system, because the medium is dielectric (optical fibers are made of silica glass) and we assume that there are no free charges (or impressed currents).}

Consider a cylindrical coordinate system $(x,r,\theta)$ (review Figure~\ref{fig:basic-diagram-bent-fiber}), related to the canonical Cartesian system by $(x, y, z) = (x, r\,\cos\theta, r\,\sin\theta)$.
Given a (constant) bend radius $\rbend > 0$ and a parameter $\sfk\ge 0$ (hereinafter known as the \emph{envelope wavenumber}),
we assume the following $\theta$-dependence ansatz for the time-harmonic electromagnetic fields:
\begin{equation}
E(x,r,\theta) = e^{-\zI\sfk \rbend \theta} \sfE(x,r,\theta), \qquad
H(x,r,\theta) = e^{-\zI\sfk \rbend \theta} \sfH(x,r,\theta).
\label{eq:envelope-ansatz}
\end{equation}
\revS{Note} that the \emph{envelope} fields $\sfE$, $\sfH$, preserve the dependence on $\theta$.
We refer to this as a \emph{full envelope ansatz}. The purpose of this modification is to shift the wavenumber of the unknown electromagnetic fields, so that we just have to solve for lowly-oscillatory solutions ($\sfE$ and $\sfH$) instead of waves at their full frequency ($E$ and $H$).
Note that, upon taking the curl of $E$, we obtain
\begin{equation*}
     \nabla \times E = e^{-\zI\sfk \rbend \theta}\left(
     \nabla\times \sfE - \zI \sfk \frac{\rbend}{r}\bfe_\theta \times \sfE
     \right),
\end{equation*}
where we have used the vector calculus identity $\nabla \times (g\bff) = g \nabla \times \bff + \nabla g \times \bff $. We make use of this identity (also applied to $\nabla\times H$) along with \eqref{eq:envelope-ansatz} to transform the time-harmonic Maxwell system \eqref{eq:time-harmonic-Maxwell}.
Denoting $\bsfK:=\sfk\frac{\rbend}{r}\bfe_\theta \times$, the new \emph{envelope Maxwell} system yields:
\begin{equation}
\left\{
\begin{array}{rl}
    \nabla\times \sfE 
    - \zI\bsfK \sfE 
    + \zI\omega\bsmu\sfH &= \bzero, \\
    \nabla\times \sfH 
    - \zI\bsfK \sfH 
    - \zI\omega\bsveps\sfE &= \bzero.
\end{array}
\right.
\label{eq:envelope-Maxwell}
\end{equation}

An equivalent form of defining $\bsfK$, in terms of Cartesian coordinates instead of cylindrical coordinates, is through the skew-symmetric matrix
$$
    \bsfK=\frac{\sfk \rbend}{y^2+z^2}
    \begin{pmatrix}
        0 & -y & -z \\
        y &  0 &  0 \\
        z &  0 &  0
    \end{pmatrix}.
$$

\subsection{Construction of PML for the curved geometry}
\label{sec:pml}

The \revisedA{\emph{perfectly matched layer}, or PML,} is a finite region in the outer part of the computational domain that \revA{replaces} an open domain in a given direction and avoids the generation of reflecting waves from the artificial boundaries.

We propose the use of PMLs for two purposes: (1) to absorb the outgoing wave after truncating the domain in the longitudinal direction, which we call \emph{PML in $\theta$}; and (2) to absorb the radiation caused by the curvature \revA{using what we call} \emph{PML in $r$} (for a cylindrical shape) or \emph{PML in $\rho$} (for a toroidal shape). We next describe how to construct them.


\begin{figure}
    \centering
    \includegraphics
    [width=0.4\linewidth]
    {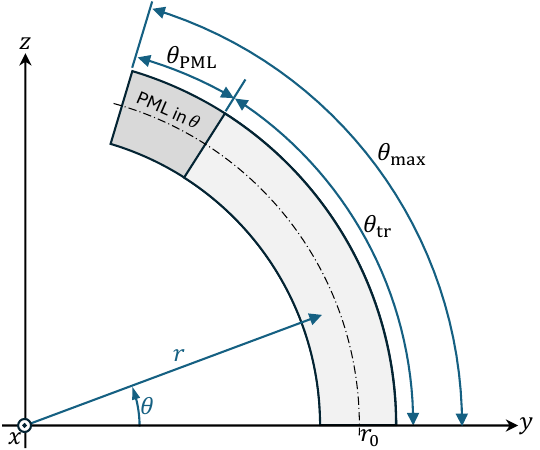}
    \caption{Circularly coiled fiber schematic that illustrates the PML at the end of the waveguide used to absorb the optical field exiting the fiber. In this case, perfect electric conductor (PEC) boundary conditions are \revS{imposed on the boundaries transverse to the circumferential direction of propagation}.}
    \label{fig:pml_theta_diagram}
\end{figure}

To impose a decaying behavior in the circumferential direction,
we propose an angle complex stretching $\theta\mapsto \tilde{\theta}$, with 
\begin{equation}
    \tilde{\theta} \defeq \tilde{\theta}(\theta,\kappa) = 
    \begin{cases}
        \theta & 0 \leq \theta \leq \theta_{\text{tr}} \, , \\
        \theta - \zI \dfrac{\pmlparameter}{\kappa} \left( \dfrac{ \theta - \theta_{\text{tr}} }{ \theta_{\text{PML}} } \right)^m
        & \theta_{\text{tr}} < \theta \leq \theta_{\text{tr}} + \theta_{\text{PML}} \, ,
    \end{cases}
    \label{eq:pml-transformation}
\end{equation}
where $\pmlparameter$ is known as the PML strength parameter, $\kappa$ is a characteristic wavenumber or frequency, $\theta_{\text{tr}}$ is the transition point at which $\theta$ enters the PML, $\theta_{\text{PML}}$ is the PML length, and the power $m\ge 2$ is a natural number. These parameters are specified by the user for each problem.

The other orthogonal directions can be complex-stretched for the sake of enforcing absorbing \revS{boundary conditions} in these directions. 
For example, the complex stretchings in $\tilde{r} = \tilde{r}(r, \kappa)$ and $\tilde{x} = \tilde{x}(x, \kappa)$ are each formulated analogously to~\eqref{eq:pml-transformation}. 
Later, we will implement the PML in the outward radial direction since the waveguide bending is expected to result in unguided radiation traveling in this same direction. 
\subsubsection{PML Jacobian}

Let $\tilde{J}_\text{cyl}$ be the uniaxial stretching Jacobian 
expressed in the local cylindrical basis, and $Q_\theta$ be the rotation matrix from Cartesian components to \revisedA{c}ylindrical components:
$$
\tilde{J}_\text{cyl} = 
    \begin{pmatrix}
        \pdv{\tilde{x}}{x} & 0 & 0 \\
        0 & \pdv{\tilde{r}}{r} & 0 \\
        0 & 0 & \pdv{\tilde{\theta}}{\theta}
    \end{pmatrix},
    \qquad
    Q_\theta = 
    \begin{pmatrix}
        1 & 0 & 0 \\
        0 & \cos\theta & \sin\theta \\
        0 & -\sin\theta & \cos\theta
    \end{pmatrix} ,
$$
Then, define $J_\text{cyl}=Q_\theta^\T \tilde{J}_\text{cyl} Q_\theta$. 
We use $J_\text{cyl}$ to \revS{express} the effects of the cylindrical coordinate stretching \revS{on} vector fields \revS{represented} in Cartesian components. \revS{Note} that the determinant is 
$|J_\text{cyl}| = \pdv{\tilde{x}}{x} \pdv{\tilde{r}}{r} \pdv{\tilde{\theta}}{\theta}$. 
\subsubsection{PML for the toroidal geometry}

The \revS{toroidal geometry is described by the} mapping from cylindrical coordinates $(x, r, \theta) \mapsto (\varphi, \rho, \theta)$:
$$
    \left\{
    \begin{array}{l}
        \varphi = \arctan{\frac{x}{r - \rbend}} \\
        \rho    = \sqrt{x^2 + (r - \rbend)^2} \\
        \theta  = \theta
    \end{array}
    \right.
$$
which generates the rotation matrix
$$
    Q_{\varphi} = 
        \begin{pmatrix}
            \sin\varphi & \cos\varphi   & 0 \\
            \cos\varphi & -\sin\varphi  & 0 \\
                      0 &           0   & 1 \\
        \end{pmatrix}.
$$
With the stretch map $(\varphi, \rho, \theta) \mapsto \big( \varphi, \tilde{\rho}(\rho, \kappa), \tilde{\theta}(\theta, \kappa) \big)$, the Jacobian in the local basis is 
$$
    \tilde{J}_\text{tor}=
        \begin{pmatrix}
            1 & 0 & 0 \\
            0 & \pdv{\tilde{\rho}}{\rho} & 0 \\
            0 & 0 & \pdv{\tilde{\theta}}{\theta}
        \end{pmatrix}
$$
making the full Jacobian become: 
$J_\text{tor} = Q_{\theta}^{\T} Q_{\varphi}^{\T} \tilde{J}_{\text{tor}} Q_{\varphi} Q_{\theta}$, and its determinant $|J_\text{tor}| = \pdv{\tilde{\rho}}{\rho}\pdv{\tilde{\theta}}{\theta}$. 

\subsubsection{Pull-back of the envelope Maxwell model}
The modified Maxwell system with the stretched coordinates reads:
\begin{equation}
\left\{
\begin{array}{rl}
    \tilde{\nabla}\times \tilde{\sfE} 
    - \zI\bsfK \tilde{\sfE} 
    + \zI\omega\bsmu\tilde{\sfH} &= \bzero, \\
    \tilde{\nabla}\times \tilde{\sfH} 
    - \zI\bsfK \tilde{\sfH} 
    - \zI\omega\bsveps\tilde{\sfE} &= \bzero.
\end{array}
\right. 
\label{eq:stretched-envelope-Maxwell}
\end{equation}

Let $J$ represent any of the PML Jacobians described above, either $J_\text{cyl}$ or $J_\text{tor}$. In order to pull-back the envelope Maxwell system from the (complex) stretched coordinates to the physical coordinates, we must apply Piola maps to $\tilde{\sfE}$, $\ \tilde{\sfH}$, $\ \tilde{\nabla}\times\tilde{\sfE}$ and $ \tilde{\nabla}\times\tilde{\sfH}$. The map for $\tilde{\sfE}$, $\tilde{\sfH}$ is the usual one for $H(\curl,\cdot)$ fields, while for their curls we use the Piola map for $H(\div,\cdot)$ fields, namely:
$$
\tilde{\sfE}=J^{-\T}\sfE, \qquad 
\tilde{\sfH}=J^{-\T}\sfH, \qquad
\tilde{\nabla}\times\tilde{\sfE}=\frac{1}{|J|}J\nabla\times\sfE, \qquad
\tilde{\nabla}\times\tilde{\sfH}=\frac{1}{|J|}J\nabla\times\sfH.
$$
After introducing these transformations into \eqref{eq:stretched-envelope-Maxwell} and multiplying both equations by an appropriate factor (${|J|J^{-1}}$) we get the final pulled-back system:

%
%
\begin{align}
    \mbox{} & \left\{
        \begin{array}{rl}
            \nabla\times \sfE - \zI \underbrace{{|J|J^{-1}}\bsfK {J^{-\sfT}}}_{\bsfK_{\text{PML}}}\sfE 
            + \zI \omega \underbrace{{|J|J^{-1}} \bsmu {J^{-\sfT}}}_{\bsmu_{\text{PML}}}\sfH &= 0,  \\
            \nabla\times \sfH - \zI \underbrace{{|J|J^{-1}}\bsfK {J^{-\sfT}}}_{\bsfK_{\text{PML}}}\sfH 
            - \zI \omega \underbrace{{|J|J^{-1}}\bsveps {J^{-\sfT}}}_{\bsveps_{\text{PML}}}\sfE &=  0;
        \end{array}
    \right. \notag \\
    \mbox{} & \Updownarrow \notag \\
    \mbox{} & \left\{
        \begin{array}{rl}
            \nabla\times \sfE - \zI \bsfK_{\PML}\sfE + \zI \omega \bsmu_{\PML}  \sfH &= 0,  \\
            \nabla\times \sfH - \zI \bsfK_{\PML}\sfH - \zI \omega \bsveps_{\PML}\sfE &= 0.
        \end{array}
    \right.
    \label{eq:PML-envelope-Maxwell}
\end{align}

In \cite{henneking2025envelope}, since the envelope ansatz dependence coincides with the PML stretched coordinate, certain simplifications are feasible within the rotation terms ($\bsfK_{\PML}\tilde{\sfE}$ and $\bsfK_{\PML}\tilde{\sfH}$) but, for the PML in the \revA{directions transverse to} the optical field propagation, those simplifications are not applicable. 
For that reason, we leave the system in its more general form \eqref{eq:PML-envelope-Maxwell}. 

From this point onward, we assume that when tensors $\bsveps$, $\bsmu$ and $\bsfK$ appear in the equations, they already include the PML transformations defined in \eqref{eq:PML-envelope-Maxwell}, omitting the specific subscript.

\subsection{Definition of functional spaces and boundary operators}
\revisedB{We assume that the domain $\Omega$ is bounded, with its  boundary $\bdry\Omega$ partitioned into a finite number of piecewise smooth subsets $\{\Gamma_m ; m=1,2,\dots,M\}$.
In general,} we are working with the space of complex-valued vector fields that are square-Lebesgue-integrable over $\Omega$, $L^2(\Omega;\C^3)$. Its inner product is denoted by $(\cdot,\cdot)_\Omega$ and its norm $\|\cdot\|_\Omega$ is \revisedB{defined by
$$
(\sfE,\sfF)_{\Omega}:=\int_\Omega \sfF^* \sfE\;dx, \qquad
\| \sfF \|_{\Omega} = \sqrt{(\sfF,\sfF)_{\Omega}}.
$$
}%
When acting on groups of vector fields, say $(\sfE,\sfH)^\T$ and $(\sfF,\sfG)^\T$, the inner product acts by pairing as follows:
$$
\left((\sfE,\sfH)^\T,(\sfF,\sfG)^\T\right)_\Omega = (\sfE,\sfF)_\Omega + (\sfH,\sfG)_\Omega\ \revisedA{.}
$$
\revisedB{
We now introduce an additional Sobolev space relevant for Maxwell's equations:
$$
H(\curl,\Omega):= \{ F\in L^2(\Omega;\C^3):\ \nabla\times F \in L^2(\Omega;\C^3)\}.
$$
%
Let us define the tangential trace operator for smooth fields, $\gamma^{\bdry\Omega}_{t}: C^\infty(\overline{\Omega};\C^3)\to C^\infty(\bdry\Omega;\C^3)$, by 
$\revisedB{\gamma^{\bdry\Omega}_{t}}\bff:=\nml\times \bff|_{\bdry\Omega} \times\nml$, where $\nml$ is the unit outward normal vector of $\bdry\Omega$. 
It is known that, for a polyhedral or piecewise smooth boundary, $\gamma^{\bdry\Omega}_t$ can be extended to a well-defined, surjective, continuous linear operator mapping $H(\curl,\Omega)$ onto $H^{-1/2}(\curl,\bdry\Omega)$ (see \cite[\S 4]{demkowicz2018energy} or \cite[\S 3.3]{buffa2001traces}). The existence of this trace operator lets us define a new subspace of $H(\curl,\Omega)$ with homogeneous boundary conditions on a subset of the boundary
$\Gamma_m\subseteq\bdry\Omega$:
$$
H_{\Gamma_m}(\curl,\Omega):= \{ F\in H(\curl,\Omega):\ \gamma^{\bdry\Omega}_t (F)|_{\Gamma_m} = 0 \}.
$$
Moreover, by $H^{-1/2}(\curl,\Gamma_m)$ we understand the space consisting of (distributional) vector fields in $H^{-1/2}(\curl,\bdry\Omega)$ restricted to the boundary portion $\Gamma_m$. This definition induces a trace operator $\gamma_t^{\Gamma_m}:H(\curl,\Omega)\to H^{-1/2}(\curl,\Gamma_m)$, defined as the composition of $\gamma_t^{\bdry\Omega}$ with the restriction to ${\Gamma_m}$.
For additional non-trivial details concerning the definition of the boundary spaces, we refer to \cite{demkowicz2018energy}.
Below, the parameters $\Omega$, $\bdry\Omega$ or $\Gamma_m$ in these spaces can be substituted by any other subdomain, its boundary, or a part of the latter, respectively.
}

\subsection{Strong form of the BVP}
From \eqref{eq:envelope-Maxwell}, we can identify the differential operator $\sfA$ acting on the unknowns $\sfu=(\sfE,\sfH)^\T$:
\begin{equation}
    \sfA\sfu \coloneqq 
    \begin{pmatrix}
    -\zI\omega\bsveps\sfE    + \nabla\times \sfH -\zI\bsfK \sfH \\
    \nabla\times \sfE -\zI\bsfK\sfE + \zI\omega\bsmu\sfH  
    \end{pmatrix}
    =
    \begin{pmatrix}
    -\zI\omega\bsveps    & -\zI\bsfK + \nabla\times \\
     -\zI\bsfK + \nabla\times  & \zI\omega\bsmu
    \end{pmatrix}
    \begin{pmatrix}
    \sfE \\ \sfH    
    \end{pmatrix}.
    \label{eq:opA}
\end{equation}

Now, suppose that the problem being analyzed imposes essential (Dirichlet) non-homogeneous linear boundary conditions on the tangential trace of $\sfE$ or $\sfH$ on $\Gamma_{E},\Gamma_{H}\subseteq\bdry\Omega$ (with $\overline{\Gamma_{E}\cup\Gamma_{H}}=\bdry\Omega\ \wedge\ \Gamma_{E}\cap\Gamma_{H}=\emptyset$), respectively. 
\revisedA{As a consequence, if the boundary condition data $\sft_\bc$ belong to the space $H^{-1/2}(\curl,\Gamma_E)\times H^{-1/2}(\curl,\Gamma_H)$, then by the surjectivity of the tangential trace operator, hence of the boundary operator
$$
\sfT=\left(\begin{matrix} \gamma^{\Gamma_E}_t & 0 \\ 0 & \gamma^{\Gamma_H}_t \end{matrix}\right): 
H(\curl,\Omega) \times H(\curl,\Omega) \longrightarrow
H^{-1/2}(\curl,\Gamma_E)\times H^{-1/2}(\curl,\Gamma_H)
,
$$ 
there exist pre-images for $\sft_\bc$ in $[H(\curl,\Omega)]^2$. Any such pre-images are candidates for $(\sfE_\lift,\sfH_\lift)^\T$. In finite element computations, boundary condition (b.c.) data are first interpolated on the boundary and then lifted with finite element shape functions to deliver the smallest support possible.
We use the notation $\sfT\sfu=\sft_{\bc}$ to represent the boundary operator acting on the unknowns of our problem.}

We can comply with the prescribed boundary condition through the corresponding conforming lifts $\sfu_{\lift}=(\sfE_{\lift},\sfH_{\lift})$; that is, 
any \revA{pair} of vector fields that belong to the graph space of $\sfA$, namely,
$$
\left\{
\sfu=(\sfE,\sfH)^\T\in[L^2(\Omega;\C^3)]^2
\ \text{s.t. }
\sfA\sfu\in[L^2(\Omega;\C^3)]^2
\right\},
$$
for which $\sfT\sfu_\lift=\sft_\bc$. 

Now that the lifts are taking care of the possible non-homogeneous traces, we can restrict the domain of $\sfA$ to the vector space of fields in the graph space that also satisfy the homogeneous version of the desired boundary condition:
$$
D(\sfA)=
\left\{
\sfu=(\sfE,\sfH)^\T\in[L^2(\Omega;\C^3)]^2
\ \text{s.t. }
\sfA\sfu\in[L^2(\Omega;\C^3)]^2
\ \text{and }
\sfT\sfu=0
\right\}.
$$
\revisedA{Note that, according to the definitions above, 
$D(\sfA) = H_{\Gamma_E}(\curl,\Omega) \times H_{\Gamma_H}(\curl,\Omega)$.}

The solution to the BVP is then an element of the affine space $\sfu_\lift+D(\sfA)$. To find the appropriate element of $D(\sfA)$, we state the strong problem as follows:
\begin{equation}
    \left\{ 
    \begin{array}{l}
    \text{Find }\sfu\in D(\sfA)\text{ such that} \\
    (\sfA\sfu,\sfv)_\Omega=(-\sfA\sfu_{\lift},\sfv)_\Omega 
    \ \text{ for every } \sfv\in [L^2(\Omega;\C^3)]^2.
    \end{array}
    \right.
    \label{eq:strong_abstract}
\end{equation}

The right-hand-side operator of \eqref{eq:strong_abstract}, \revisedB{which involves the lift of the Dirichlet data}, will be included in the continuous conjugate-linear functional $\ell$ in the variational formulation \revisedB{detailed below.}

The well-posedness (in particular, the verification of the inf--sup condition in the continuous setting) for the strong form of the envelope Maxwell boundary value problem follows from that of the classical Maxwell system, thanks to the invariance of the norm under $e^{-\zI\sfk \rbend\theta}$ (for details, see subsection \emph{Waveguide stability analysis} in \cite{henneking2025envelope}).

\subsection{Ultraweak formulation}
The (formal) adjoint of $\sfA$, acting on test functions $\sfv=(\sfF,\sfG)^\T$, corresponds to
\begin{equation}
    \sfA^*\sfv =
    \begin{pmatrix}
    -\overline{\zI\omega\bsveps^\T}    & -\overline{\zI\bsfK^\T} + \nabla \times \\
    -\overline{\zI\bsfK^\T} + \nabla \times  & \overline{\zI\omega\bsmu^\T}
    \end{pmatrix}
    \begin{pmatrix}
    \sfF \\ \sfG    
    \end{pmatrix}
    =
    \begin{pmatrix}
    -\overline{\zI\omega\bsveps^\T}\sfF + \nabla\times\sfG -\overline{\zI\bsfK^\T} \sfG \\
    \nabla\times \sfF -\overline{\zI\bsfK^\T}\sfF + \overline{\zI\omega\bsmu^\T}\sfG  
    \end{pmatrix}
    .
    \label{eq:opA*}
\end{equation}
Because we need to enforce $(\sfA\sfu,\sfv)_\Omega=(\sfu,\sfA^*\sfv)_\Omega$ for every $\sfu\in D(\sfA)$ and $\sfv\in D(\sfA^*)$,
a complete definition of the adjoint requires determining its domain as well. 
Since $D(\sfA)$ includes the observance of certain boundary conditions, so should $D(\sfA^*)$, hence
$$
D(\sfA^*)=
\left\{
\sfv=(\sfF,\sfG)^\T\in[L^2(\Omega;\C^3)]^2
\ \text{s.t. }
\sfA^*\sfv\in[L^2(\Omega;\C^3)]^2
\ \text{and }
\sfT^*\sfv=0
\right\}.
$$
Here, $\sfT^*$ is not an adjoint per se, but it stands for a possible change with respect to $\sfT$. \revisedA{We prefer not to go deeper into this new boundary operator because in the broken ultraweak formulation, which we ultimately implement, there is no need for it.}

Having finished these definitions, take the trial space $\scU_0=[L^2(\Omega;\C^3)]^2$, and the test space $\scV_0=D(\sfA^*)$ equipped with the topology induced by the adjoint-graph inner product,
$$
(\delta{\sfv},\sfv)_{\scV_0} = (\sfA^*\delta{\sfv} , \sfA^*\sfv)_\Omega + \alpha (\delta{\sfv},\sfv)_\Omega,
$$
for some positive constant $\alpha$. Consider any continuous conjugate-linear functional $\ell\in\scV_0'$ \revisedB{acting on the elements $\sfv$ of the test space. This load is included in this form for the sake of generality, even though in our context, as mentioned above, $\ell$ consists only of the boundary data lift (see the right-hand side of \eqref{eq:strong_abstract}).
The ultraweak variational formulation reads}
\begin{equation}
    \left\{ 
    \begin{array}{l}
    \text{Find }\sfu\in\scU_0\text{ such that} \\
    (\sfu,\sfA^*\sfv)_\Omega = \ell(\sfv) \ \text{ for every } \sfv\in\scV_0.
    \end{array}
    \right.
    \label{eq:uw_abstract}
\end{equation}
The left-hand-side sesquilinear form in \eqref{eq:uw_abstract} is next expanded:
\begin{equation}
    (\sfu,\sfA^*\sfv)_\Omega 
    =(\sfE, \nabla\times\sfG -\overline{\zI\bsfK^\T} \sfG -\overline{\zI\omega\bsveps^\T}\sfF  )_\Omega 
    + (\sfH,\nabla\times \sfF -\overline{\zI\bsfK^\T}\sfF + \overline{\zI\omega\bsmu^\T}\sfG)_\Omega.
\label{eq:uw_lhs}
\end{equation}

In general, the well-posedness of an ultraweak formulation holds if and only if the respective strong formulation is well-posed. That deep result is based on arguments related to Banach's closed range theorem for closed operators, and is explained in detail in multiple prior DPG publications such as, for instance, the recent review \cite[\S 7]{Demkowicz_Gopalakrishnan_2025}. Since the well-posedness of the strong version has already been established for this problem, the well-posedness of this ultraweak form is then automatically observed.

%% file: 3_discretization.tex
\section{Discretization and implementation}
\label{sec:discretization}

Now that the ultraweak formulation is established, we can propose its numerical approximation through the discontinuous Petrov--Galerkin (DPG) method. To do so, we start by incorporating broken test spaces into the formulation. This is followed by the description of the practical application of DPG to the present setting, and we close this section with a brief summary of our computational implementation's details.

\subsection{Formulation with broken test spaces}

Given a regular mesh or triangulation $\mesh$ of the domain $\Omega$, \revisedB{where each element $K\in\mesh$ is itself a bounded Lipschitz domain with piecewise smooth boundary, let us introduce the broken (or product) space
$$
H(\curl,\mesh):=\{F\in L^2(\Omega;\C^3): F|_K \in H(\curl,K)\}
\cong
\prod_{K\in\mesh}H(\curl,K),
$$
and the space for the interface/trace fields defined on the mesh skeleton $\bdry\mesh$,
$$
H^{-1/2}(\curl,\bdry\mesh):=\left\{\prod\limits_{K\in\mesh}\gamma^{\bdry K}_t (F|_K): F\in H(\curl,\Omega) \right\} .
$$
}

Now, consider the broken version of $\sfA^*$, namely,
$$
\sfA^*_h\sfv :=
\begin{pmatrix}
-\overline{\zI\omega\bsveps^\T}    & -\overline{\zI\bsfK^\T} + \nabla_h \times \\
-\overline{\zI\bsfK^\T} + \nabla_h \times  & \overline{\zI\omega\bsmu^\T}
\end{pmatrix}
\begin{pmatrix}
\sfF \\ \sfG    
\end{pmatrix},
$$
where the curl operator $\nabla_h\times$\, is applied elementwise,
and the broken test space $\scV=[H(\curl,\mesh)]^2$, \revS{on which no essential boundary condition are imposed}. Testing with these broken spaces induces the appearance of two new unknowns: $\check{\sfu}=(\check{\sfE},\check{\sfH})$ from
$\check{\scU}=\{\check{\sfE},\check{\sfH} \in H^{-1/2}(\curl,\bdry\mesh) \text{ such that } \check{\sfE},\check{\sfH}\text{ satisfy essential boundary conditions}\}$. We refer to \revB{these} as the \emph{interface unknowns} or \emph{traces}. Their introduction into the formulation is through a new term in the sesquilinear form:
$$
    \langle \check{\sfu}, \sfv \rangle_{\partial\mesh} \defeq  
        \sum\limits_{K \in \mesh} 
        \langle \nml \times \check{\sfE} , \revisedB{\gamma^{\bdry K}_t}(\sfG|_{K}) \rangle_{\bdry K} +
        \langle \nml \times \check{\sfH} , \revisedB{\gamma^{\bdry K}_t}(\sfF|_{K}) \rangle_{\bdry K} \ ,
$$
understanding $\langle\cdot,\cdot\rangle_{\bdry K}$ as the duality pairing between $H^{-1/2}(\div,\bdry K)$ and 
$H^{-1/2}(\curl,\bdry K)$\footnote{\revisedA{In fact, the space $H^{-1/2}(\div,\bdry K)$ is precisely defined as the dual to $H^{-1/2}(\curl,\bdry K)$.}}.

The updated problem with broken test spaces reads:
\begin{equation*}
    \left\{ 
    \begin{array}{l}
        \text{Find }(\sfu,\check{\sfu})\in\scU:=\scU_0\times\check{\scU}\text{ such that} \\[6pt]
        \underbrace{(\sfu,\sfA_h^*\sfv)_\Omega + \langle \check{\sfu}, \sfv \rangle_{\partial\mesh}}_{b((\sfu,\check{\sfu}),\sfv)}
        = 
        \ell(\sfv) \ \text{ for every } \sfv\in\scV.
    \end{array}
    \right.
\end{equation*}
The adjoint-graph inner product must also be adjusted: 
$(\delta{\sfv}, \sfv)_{\scV} = (\sfA_h^*\delta{\sfv} , \sfA_h^*\sfv)_\Omega + \alpha (\delta{\sfv}, \sfv)_\Omega$.

\subsection{Applying the practical DPG method}

By choosing a conforming trial discrete space $\scU_h\subset\scU$ along with a finite-dimensional enriched test space $\scV_r\subset\scV$ ($\dim \scV_r > \dim \scU_h$), the practical DPG method seeks the discrete solution $(\sfu_h,\check{\sfu}_h)\in \scU_h$ as the minimizer of
$$
    \| \ell(\cdot) - b((\sfu_h,\check{\sfu}_h),\cdot) \|_{\scV_r'}^2 \;,
$$
with the dual norm induced by the inner product assigned to $\scV$ above.
To solve it, the residual minimization problem is restated as a saddle point problem:

Find $\psi_r\in \scV_r$ and $(\sfu_h, \check{\sfu}_h) \in \scU_h$ such that 
\begin{equation}
    \left\{
    \begin{array}{ccll}
         (\psi_r,\sfv_r)_\scV &+\quad b((\sfu_h,\check{\sfu}_h),\sfv_r) & =\ell(\sfv_r) & \text{for every }\sfv_r\in\scV_r,   \\
         b^*(\psi_r,(\delta\sfu,\delta\check{\sfu})) &   & = 0           & \text{for every }(\delta\sfu,\delta\check{\sfu})\in\scU_h.
    \end{array}
    \right.
    \label{eq:mixed_dpg}
\end{equation}

In this discrete mixed formulation, the unknowns are
\begin{itemize}
    \item \revS{The residual representation ${\psi_r=(\sfF_{\psi_r},\sfG_{\psi_r})^\T}$, consisting of two broken $H(\curl,\mesh)$ vector fields that are each discretized with} discontinuous piecewise N\'{e}d\'{e}lec polynomials of degree $p+dp$, \revisedA{where $p\in\Z^+$ is the order of approximation and $dp\in\Z^+$ is the enrichment order for the test functions}.
    \item Two $L^2(\Omega)$ vector fields $\sfu_h=(\sfE_h,\sfH_h)^\T$, discretized with discontinuous piecewise (tensor-product) polynomials of degree $p-1$.
    \item Two tangential-trace $H^{-1/2}(\curl,\partial\mesh)$ vector fields $\check{\sfu}_h=(\check{\sfE}_h,\,\check{\sfH}_h)^\T$, discretized with tangential traces (over the mesh skeleton) of conforming N\'{e}d\'{e}lec polynomials of degree $p$.
\end{itemize}

The test functions are
$\sfv_r=(\sfF_r,\sfG_r)^\T,\, {\delta\sfu_h=(\delta\sfE_h,\delta\sfH_h)^\T,\,\delta\check{\sfu}_h=(\delta\check{\sfE}_h,\,\delta\check{\sfH}_h)^\T}$, discretized with the same spaces \revS{described} above.

\revisedA{The choice of element spaces, in the case of trial functions, is based on N\'{e}d\'{e}lec's polynomial exact sequence for the hexahedron of the first kind with order of approximation $p$. Regarding the degree for the enriched test functions $\sfv_r$ (and the residual representation $\psi_r$), we are applying the minimum enrichment ($dp=1$) based on numerical evidence for hexahedral elements. 
Results on the existence of Fortin operators for tetrahedral elements, available in the literature \cite{BrokenForms15,demkowicz2020construction}, call for an enrichment of $dp \geq 2$.}

Given specific bases for the chosen discrete spaces, \revS{suppose that} the coefficient vector corresponding to $\psi_r$ is $\bfr$, while the \revS{coefficient vector} corresponding to $(\sfu_h, \check{\sfu}_h)$ is $\bfd$. Then, the saddle point discrete problem in matrix notation reads
\begin{equation}
\begin{bmatrix}
    \bfG   & \bfB \\
    \bfB^* & \bzero
\end{bmatrix}
\begin{bmatrix}
    \bfr\\
    \bfd
\end{bmatrix}
=
\bfl ,
\label{eq:matrix_mixed_dpg}
\end{equation}
where we have the block-diagonal Gram matrix $\bfG$, with entries $[\bfG]_{ij}=(\sfv_j,\sfv_i)_{\scV}$; the enriched stiffness matrix $\bfB$, with entries $\bfB_{ij}=b((\sfu_j,\check{\sfu}_j),\sfv_i)$; and the enriched load vector $\bfl$, with entries $\bfl_i=\ell(\sfv_i)$. Note that this algebraic mixed problem can be easily reduced to the smaller Hermitian positive-definite system
$(\bfB^*\bfG^{-1}\bfB)\;\bfd = (\bfB^*\bfG^{-1}\bfl)$, solving for $\bfd$ only. If DPG is implemented in this way, then the residual representation function $\psi_r$ may be retrieved \emph{a posteriori}, with $\bfr=\bfG^{-1}(\bfl-\bfB\bfd)$, so that its $\scV$-norm is given by:
$$
\| \psi_r \|_{\scV}^2 = \bfr ^* \bfG \bfr = (\bfl-\bfB\bfd)^* \bfG^{-1} (\bfl-\bfB\bfd).
$$
The discontinuous nature of the broken test functions, and consequently of the residual representation function, makes it possible to compute $\| \psi_r \|_{\scV}^2$ elementwise (facilitating an efficient parallel computation), since $\| \psi_r \|_{\scV}^2=\sum\limits_{K\in\mesh} \| \psi_r|_K \|_{\scV}^2$. The value $\| \psi_r|_K \|_{\scV}$ is utilized as an \emph{a posteriori} error indicator to automatically perform adaptive refinement via a greedy algorithm (we mark for refinement all elements whose local residual is at least 50\% of the maximum element residual).
Notice that, when implementing the residual computation, there is no need to explicitly form \revS{the} matrix $\bfB$, but rather we can integrate the vector $[\bfB\bfd]_i=b((\sfu_h,\check{\sfu}_h),\sfv_i)$, utilizing the previously found coefficient vector $\bfd$ (that is, $(\sfu_h,\check{\sfu}_h)$).

It is a well-established result that solving the DPG problem in this form is equivalent to solving the discrete ultraweak formulation with a set of optimal test functions (which depend on the discrete trial space, the operator and the test norm). For a deeper discussion of DPG fundamentals, we refer to \cite{demkowicz2015encyclopedia,Demkowicz_Gopalakrishnan_2025} and
\cite[{\S}5]{Demkowicz_FEbook}.

\subsection{Implementation details}

We have used the $hp\mathrm{3D}$ finite element library \cite{henneking2024hp3d}, which supports elements of all the common 3D shapes (tetrahedron, prism, hexahedron, pyramid), the energy spaces ($H^1(\cdot)$, $H(\curl,\cdot)$, $H(\div,\cdot)$, $L^2(\cdot)$), automatic adaptive local refinements, 1-irregular meshes, and hybrid MPI+OpenMP parallelization, among other advanced features. The initial meshes are designed to be interpreted by the built-in geometry package, and they can be as coarse as possible since we choose to work with exact geometry elements (e.g., cylindrical and toroidal \revS{elements} are parameterized with trigonometric functions, not with polynomials).

In order to reduce \revS{the global linear system to one involving only the trace unknowns} $\check{\sfE}_h$ and $\check{\sfH}_h$ only, we \revS{use} the static condensation capabilities of $hp\mathrm{3D}$. The fields $\sfE_h, \sfH_h$ are later retrieved locally.

The distributed-memory simulations were computed on \revS{the} Texas Advanced Computing Center's (TACC's) \emph{Frontera} supercomputer \cite{stanzione2020frontera}, using up to 128 Intel Cascade Lake compute nodes of 56 cores each (7,168 cores in total). The adaptive simulations reported in numerical experiments 2 and 3 apply dynamic load-balancing strategies \revS{based on the ParMETIS graph-partitioning library}, performed after each automatic adaptive refinement, to minimize the communication costs for the parallel solves performed with the MUMPS direct solver.

%% file: 4_experiments.tex
\section{Numerical experiments}
\label{sec:experiments}

In order to verify and demonstrate our methodology, we present three numerical experiments, of increasing complexity.

The first problem---a vacuum waveguide with \revisedB{a perfect electric conductor (PEC)} boundary---is posed essentially in a two-dimensional setting, but we implement it in our 3D code, eliminating any dependence on the additional coordinate. The exact solution can be obtained via separation of variables, so it serves as a means to verify the convergence of our discretization with different orders of approximation. This setup requires the PML only in $\theta$. Here, we perform uniform $h$-refinements and observe the behavior of the DPG residual in comparison with the numerical error to verify that they converge as expected from theory.

The second case---a step-index open slab waveguide---depends on two dimensions as well. This problem \revS{includes} a radiation condition, modeled by a PML in the radial direction. We show convergence of the DPG residual as the mesh is adaptively refined, and verify that the computed radiation losses converge to the analytically estimated losses.

The third and final case---a bent step-index optical fiber---is fully three-dimensional. Moreover, the exact solution is unknown, so we rely on the DPG error indicator for adaptive mesh refinements to accurately capture the details of the solution. We impose as a boundary condition at the input facet the modes of a straight step-index fiber, and simulate their propagation along the bent fiber domain.

A key pre-processing step for the numerical experiments is choosing appropriate physical units/scales. So, before describing the three experiments in greater detail, we show the nondimensionalization/normalization carried out, and present a set of physically realistic parameters that have been used in their implementations.

\paragraph{Physical parameters and nondimensionalization.}
To reduce the wide range of orders of magnitude to which our physical parameters belong, we choose reference values for four fundamental physical quantities: length, mass, time and electrical current (see Table \ref{tab:nondimensional}), and convert all the parameters to nondimensional units. Table \ref{tab:fiber_data} lists the parameter values in SI units and in their corresponding nondimensional units. All numerical results are presented in these nondimensionalized units. 
\begin{table}[htb]
    \centering
    \caption{Reference values used for the nondimensionalization of the numerical experiments.
    }
    \label{tab:nondimensional}
    \begin{tabular}
        {|l|r|}\hline
        Reference length              & $25.4 \cdot 10^{-6}$ m \\
        Reference mass                & $10^{-22}$ kg \\
        Reference time                & $10^{-13}$ s \\
        Reference electrical current  & $400$ A \\
        \hline
    \end{tabular}
\end{table}

\begin{table}[htb]
    \centering
    \caption{Coiled waveguide parameters in both SI and nondimensional units. All the waveguide's properties are adapted from the optical step-index fiber analyzed in \cite{henneking2021fiber}.}
    \begin{tabular}{|r|c|c|c|}
        \hline
        Parameter                       & Symbol       & Value in SI units              & Nondimensional value \\
        \hline
        Bend radius range & $\rbend$        & $[0.03302\text{ m}, 0.06604\text{ m}]$               & $[1300, 2600]$  \\
        Core's radius              & $\rcore$ ($a$)        & $1.27 \cdot 10^{-5}$ m          & $0.5$ \\
        Cladding's radius            & $\rclad$ ($b$)          & $1.27 \cdot 10^{-4}$ m        & $5.0$ \\
        Fiber's (waveguide's) radius & $\rfi$          & $2.54 \cdot 10^{-4}$ m        & $10.0$ \\
        Core's refractive index        & $\ncore$       & 1.4512                         & $1.4512$\\
        Cladding's refractive index    & $\nclad$       & 1.45                            & $1.45$\\
        Coating's refractive index & $\ncoat$  & 1.38                         & 1.38\\
        Wavelength                     & $\lambda$      & $1.064 \cdot 10^{-6}$ m         & $0.0418897637795276$ \\
        Free-space wavenumber        & $k_0$        & $5.905 \cdot 10^{6}$ rad/m        & $1.49993333460866 \cdot 10^{2}$ \\
        Angular frequency           & $\omega$   & $1.77035 \cdot 10^{15}$ rad/s      & $1.77034921739554 \cdot 10^{2}$ \\
        Vacuum permeability            & $\mu_0$      & $1.25664 \cdot 10^{-6}$ N/A$^2$ & $0.791582400800000$ \\
        Vacuum permittivity            & $\epsilon_0$ & $8.85419 \cdot 10^{-12}$ F/m    & $0.906838390340768$ \\
        \hline
    \end{tabular}
    \label{tab:fiber_data}
\end{table}

\begin{remark}
    We note that both the glass and the polymer coating materials have some small intrinsic losses. For the polymer, such losses might be quantified within 3--40 dB/km~\cite{tankala2011reliability, coherent2026nucoat}, and, for glass, the losses are even smaller. When interpreting this attenuation factor as the imaginary part of the refractive index of the coating, the result is at most of order $10^{-9}$. On the other hand, the real part of the refractive index is of order 1. Because of this enormous contrast, and because it does not alter our methodology for computing confinement losses, we neglect these intrinsic losses (i.e., assume that the imaginary parts of ${\ncore}$, ${\nclad}$, and ${\ncoat}$ are all zero), and we furthermore set the refractive index of air equal to that of vacuum, ${\nair} = 1$~\cite{ciddor1996refractive}.
\end{remark}

\paragraph{PML design and parameters.}
\revA{
The location and type of the perfectly matched layers used in each numerical experiment are determined by the geometry and the expected direction of outgoing radiation. 
All three examples require a PML in the $\theta$ direction because the computational domain is truncated to a short arc, while the optical field should behave as if it continued to propagate along the bent waveguide without reflection from the artificial output boundary. 
The treatment of the transverse boundaries differs among the three experiments. 
In the first experiment, the transverse boundaries are perfectly conducting (PEC boundaries), so no radial PML is required. 
In the second and third experiments, the transverse domain is intended to represent an open waveguide, with the cladding in experiment~2 and the coating in experiment~3 effectively extending beyond the computational boundary.
}

\revA{
The asymmetric radial treatment in experiment~2, where a PML is applied only at the outer radial boundary, is motivated by both the physics of bend radiation and our previous numerical observations. 
Classical analyses of bending loss describe the radiating field on the outer side of the bend through asymptotic Hankel-function behavior as $r\to\infty$, while the field on the inner side is represented by a large-order Bessel function of the first kind and decays exponentially as $r$ decreases away from the core \cite{marcatili1969bends, marcuse1982light}. 
Such bending-loss models have been shown to agree well with experimental measurements \cite{schermer2007improved}.
Consistent with these observations, our numerical study in \cite{mora2025bessel} showed that the propagating-mode profiles (i.e., the eigenfunctions) are only weakly sensitive to the precise location and type of boundary condition imposed on the inner radial boundary, whereas the radiating oscillations develop on the outer side of the bend. 
We therefore use the same homogeneous Neumann b.c.\ as in \cite{mora2025bessel} on the inner boundary, imposing a vanishing magnetic field trace; this also allows the modes computed there to be used consistently for verification of the present finite element simulations.
}

\revA{
In experiment~3, transverse radiation must again be absorbed, but the toroidal fiber geometry makes it more convenient to surround the cross-section with a complete PML in the $\rho$ direction. This full annular PML simplifies the mesh construction while providing absorption of radiation leaving the fiber through the outer coating region.
}

In all cases, the PML parameters of \eqref{eq:pml-transformation} are set to $\pmlparameter = 200$ and $m=2$. However, the frequency or wavenumber involved in the complex-stretching map, $\kappa$, is taken as follows:
\begin{itemize}
    \item \textbf{Bent longitudinal direction}: in the first numerical experiment, $\kappa = (k_0 - \sfk) \rbend$; in the second and third experiments, $\kappa = (k_0 \ncore - \sfk) \rbend$.
    \item \textbf{Cross-sectional direction}: in the second numerical experiment, $\kappa = k_0 \nclad$; in the third experiment, $\kappa = k_0 \ncoat$.
\end{itemize}

The location of the transition point $\theta_{\text{tr}}$ between the inner domain and the PML in $\theta$ is set such that the corresponding PML length, $\theta_{\text{PML}}=\theta_{\max}-\theta_{\text{tr}}$, is about two \emph{envelope beat lengths}. The envelope beat length is the result of $2\pi/(k_{\text{eff}}-\sfk)$, where the effective wavenumber $k_{\text{eff}}$ is approximated with the propagation constant of the straight waveguide fundamental mode.
In the second and third numerical experiments, the lengths of the transverse PML (in $r$ or in $\rho$) are set to cover half the width of the outermost layer (i.e., the cladding or the coating, respectively). The transition points ($r_{\text{tr}}, \rho_{\text{tr}}$) are set accordingly.

\paragraph{Verification using propagating modes of a bent slab waveguide.}
We study the transverse modes for the electric field (TE modes), polarized in $x$; that is, we have $E_r=E_\theta\equiv 0$, and assume that $E_x$ is in the form
$$
E_x(x,r,\theta)=e^{-\zI\beta\theta}\mcE_x(x,r),
$$
where $\beta\in\C$ is the mode's propagation constant in $\theta$, and the function $\mcE_x(x,r)$ is the corresponding TE mode, the only nonzero component of $\mcE=(\mcE_x,0,0)^\T$. It is well known that Maxwell's equations \eqref{eq:time-harmonic-Maxwell}, under certain assumptions, lead to an eigenvalue problem for a scalar Helmholtz equation (subject to particular boundary and interface conditions). Incorporation of the above mode ansatz into the Helmholtz problem yields the following resulting equation for $\mcE_x(x,r)$:
\begin{equation}
    \pdv[2]{\mcE_x}{x}(x,r)+\pdv[2]{\mcE_x}{r}(x,r) + \frac{1}{r}\pdv{\mcE_x}{r}(x,r) + 
    \frac{1}{r^2}\left[ \big( k_0 n(x,r) r \big)^2 - \beta^2 \right]\mcE_x(x,r) = 0,
    \label{eq:TE-eigenvalue-problem}
\end{equation}
where $n(x,r)$ is the refractive index of the medium, $k_0 = 2 \pi / \lambda$ is the free-space wavenumber, and $\lambda$ is the optical field's wavelength. The eigenvalue in this problem corresponds to $\beta^2$, and the eigenfunction is any multiple of the mode profile $\mcE_x = \mcE_x(x, r)$. 
For details on how this eigenvalue problem can be solved numerically, we refer to \cite{gopalakrishnan2025helical, mora2025bessel}.

The time-harmonic magnetic field corresponding to this mode ansatz is $H = e^{-\zI\beta\theta}\mcH(x,r)$, where $\mcH$ has components
$$
\mcH_x = 0, \qquad 
\mcH_r = \frac{\beta}{\omega\mu_0 r}\mcE_x, \qquad
\mcH_\theta = \frac{1}{\zI\omega\mu_0}\pdv{\mcE_x}{r},
$$
a direct result of the Maxwell--Faraday law, $\zI\omega\mu_0 H=-\nabla\times E$.
In the first two numerical experiments, we use precomputed transverse modes of distinct bent waveguides, apply them as boundary conditions at the input facet of the waveguide, and evaluate the error and residual convergence for these propagating modes with the envelope Maxwell DPG model.

\subsection{Numerical experiment 1: Vacuum waveguide with perfectly conducting walls}\label{subsec:experiment1}

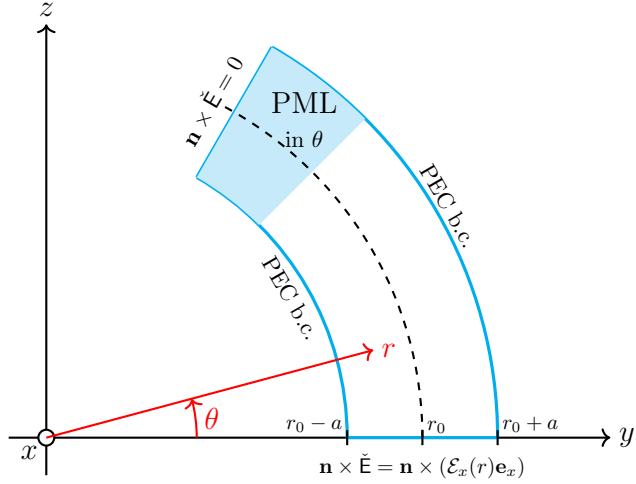
\begin{figure}
    \centering
    \begin{tikzpicture}[scale=1, thick, every node/.style={scale=0.8}]
    
    \draw[->] (-0.5,0) -- (7.5,0) node[anchor=west] {\Large $y$};
    \draw[->] (0,-0.5) -- (0,5.5) node[anchor=south] {\Large $z$};
    \draw[fill=white] (0,0) circle(3pt);
    \node[below left] at (0,0) {\Large $x$};
    
    \def\rmin{4.0}
    \def\rbnd{5.0}
    \def\rmax{6.0}
    \def\pmlangle{15}
    
    \draw[cyan, very thick] 
    ({\rmin*cos(0)}, {\rmin*sin(0)}) arc[start angle=0, end angle=45+\pmlangle, radius=\rmin];
    \draw[cyan, very thick] 
    ({\rmax*cos(0)}, {\rmax*sin(0)}) arc[start angle=0, end angle=45+\pmlangle, radius=\rmax];
    \draw[cyan, very thick] (\rmin,0) -- (\rmax,0);
    \draw[cyan, very thick] ({\rmin*cos(60)},{\rmin*sin(60)}) -- ({\rmax*cos(60)},{\rmax*sin(60)});
    
    \begin{scope}
    \clip ({\rmin*cos(45)}, {\rmin*sin(45)}) arc[start angle=45, end angle=45+\pmlangle, radius=\rmin] --
          ({\rmax*cos(45+\pmlangle)}, {\rmax*sin(45+\pmlangle)}) arc[start angle=45+\pmlangle, end angle=45, radius=\rmax] -- cycle;
    \fill[cyan!20] (0,0) circle(\rmax);
    \end{scope}
    
    \draw[dashed] 
    ({\rbnd*cos(0)}, {\rbnd*sin(0)}) arc[start angle=0, end angle=62, radius=\rbnd];
    
    \draw[->, red] (0,0) -- ({4.5*cos(15)}, {4.5*sin(15)}) node[anchor=west] {\Large $r$};
    \draw[->, red] (2.0,0) arc[start angle=0, end angle=15, radius=2.0];
    \node[red] at (2.2,0.25) {\Large $\theta$};
    
    \node[left]  at ({\rmin*cos(0)+0.05}, 0.175) {\small $\rbend - a$};
    \node[right] at ({\rbnd*cos(0)-0.05}, 0.175) {\small $\rbend\phantom{+0}$};
    \node[right] at ({\rmax*cos(0)-0.05}, 0.175) {\small $\rbend + a$};
    
    \draw[black] (\rmin,0.1) -- (\rmin,-0.1);
    \draw[black] (\rbnd,0.1) -- (\rbnd,-0.1);
    \draw[black] (\rmax,0.1) -- (\rmax,-0.1);
    
    \node at ({1.03*\rmax*cos(30)}, {1.03*\rmax*sin(30)}) [rotate=-60] {PEC b.c.};
    \node at ({0.94*\rmin*cos(30)}, {0.93*\rmin*sin(30)}) [rotate=-60] {PEC b.c.};
    \node at ({\rbnd*cos(64)}, {\rbnd*sin(64)}) [rotate=60] {$\nml\times\check{\sfE}=0$};
    \node at ({\rbnd*cos(0)}, {\rbnd*sin(0)-0.1}) [below] {\small $\bfn\times\check{\sfE} = \bfn\times(\mcE_x(r) \bfe_x)$};
    
    \node[align=center] at ({1.09*\rbnd*cos(51)}, {1.09*\rbnd*sin(51)}) {\Large PML\\ in $\theta$};

    \end{tikzpicture}
    \caption{Geometry setup and boundary conditions for numerical experiment 1: we apply \revisedB{perfect electric conductor (PEC)} b.c.\ at the curved boundaries, a PML in $\theta$, and prescribe the mode on the bottom face (waveguide input facet). The parameters defining the PML length in $\theta$ are illustrated in Figure~\ref{fig:pml_theta_diagram}.}
    \label{fig:exp1-geometry}
\end{figure}

In cylindrical coordinates $(x,r,\theta)$, the waveguide domain is $(-1/2,1/2) \times (\rbend - a, \rbend + a) \times (0, \theta_{\max})$. 
\revisedB{The domain and the setup for this example are illustrated in Figure \ref{fig:exp1-geometry}.}
We assume that the data and the solutions do not depend on coordinate $x$. The corresponding eigenvalue equation \eqref{eq:TE-eigenvalue-problem} for this setup then reduces to the Bessel eigenvalue equation:
\begin{equation}
    \odv[2]{\mcE_x}{r}(r) + \frac{1}{r}\odv{\mcE_x}{r}(r) + 
    \frac{1}{r^2}\left[ (k_0 n(r) r)^2 - \beta^2 \right]\mcE_x(r) = 0.
    \label{eq:bessel-problem}
\end{equation}
For the vacuum waveguide, we set $n(r)\equiv 1$.
At the inner and outer radial boundaries, we impose PEC boundary conditions:
$$
\bfn\times E = 0 \qquad\text{at } r = \rbend - a,\; \rbend + a.
$$
In the ultraweak DPG formulation, these PEC boundary conditions are imposed via the tangential trace unknowns, $\bfn\times \check{\sfE} = 0$.

Note that the PEC boundary condition is equivalent to imposing $\mcE_x(\rbend - a) = \mcE_x(\rbend + a) = 0$ in the reduced problem \eqref{eq:bessel-problem}. 
This Bessel eigenvalue equation, along with the given homogeneous boundary conditions, is a Sturm--Liouville problem, hence its eigenvalues are real, and the modes are orthogonal in a weighted $L^2$ space.

\begin{figure}[htb]
    \centering
    \begin{subfigure}{0.49\textwidth}
        \includegraphics[width=\textwidth]{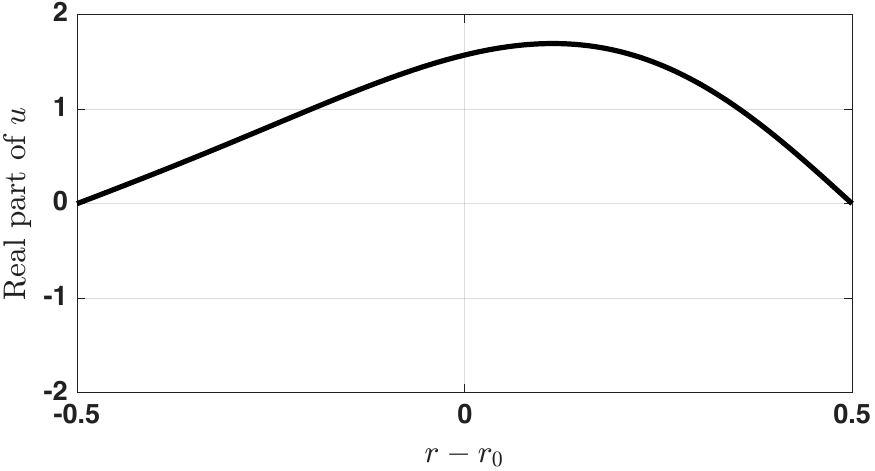}
        \caption{First mode, $\beta^2=3.8007\cdot10^{10}$}
        \label{plot:ex1-r1300-mode1-real}
    \end{subfigure}%
    \begin{subfigure}{0.49\textwidth}
        \includegraphics[width=\textwidth]{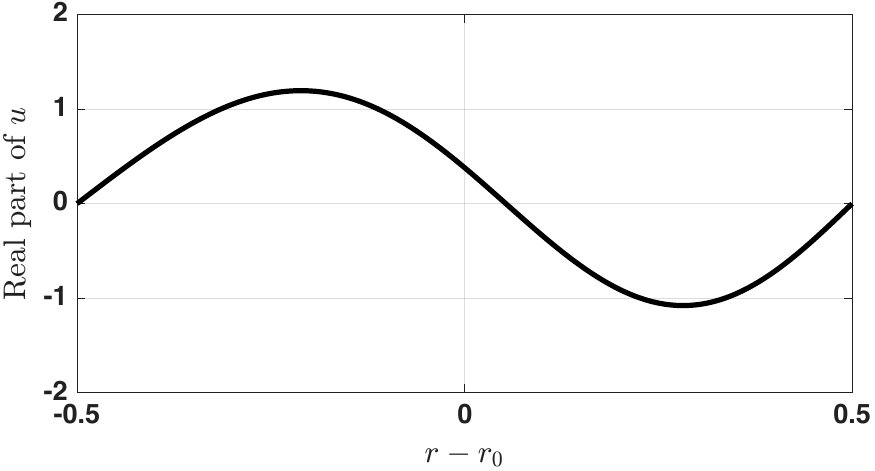}
        \caption{Second mode, $\beta^2=3.7954\cdot10^{10}$}
        \label{plot:ex1-r1300-mode2-real}
    \end{subfigure}%
    \caption{Real component of the mode profiles for the first and second modes, with corresponding propagation constant $\beta$, of the vacuum waveguide with PEC boundary, bend radius $\rbend = 1300$, and width $2a=1$.}
    \label{fig:ex1-modes-real}
\end{figure}

Every mode in this problem is a particular combination of two Bessel functions of (real) order $\beta$, such that it vanishes at both endpoints. 
The scaling condition for the mode (i.e., the eigenfunction) consists of fixing a certain value of the derivative at the left endpoint (i.e., at the inner radius $r = \rbend - a$). 
Using the high-accuracy algorithms described in \cite{mora2025bessel}, we precompute the eigenvalues and eigenfunctions for the first two modes, with bend radius $\rbend = 1300$ and width $2a=1$. Figure~\ref{fig:ex1-modes-real} depicts the profiles of these solutions, which are purely real-valued. \revS{Note} that there is a close resemblance with the first two modes of a straight rectangular waveguide of unit width, $\sin(\pi y)$ and $\sin(2\pi y)$.

These numerically computed mode solutions are then used in our 3D waveguide implementation in hp3D for imposing Dirichlet boundary conditions for the trace $\check{\sfE}$ at the waveguide input facet,
$$
    \bfn \times \check{\sfE} = \bfn \times (\mcE_x(r) \bfe_x) 
    \quad \text{at} \quad \theta = 0 ,
$$
as well as for calculating the numerical error of the 3D envelope Maxwell model solution for the propagating mode.

The desired mode is independent of coordinate $x$ (i.e., independent of the direction normal to the bending plane). This behavior is enforced by applying the b.c.\ $\bfn \times \check{\sfE} = 0$ on the back and front faces ($x = -1/2$ and $x = 1/2$, respectively). 
In this experiment, we include a PML in $\theta$ only, which is defined over the arc $(\theta_{\tr},\theta_{\max}]$.
Under the assumption that the PML makes the outgoing waves decay to a near machine-zero amplitude, we can impose $\bfn\times \check{\sfE}=0$ at the waveguide output (i.e., the top face), $\theta = \theta_{\max}$. 
For this first numerical experiment, the domain has a maximum angle of $\theta_{\max}=3^{\circ}$, with the circumferential PML starting at $\theta_{\tr}=2.25^{\circ}$, and no PML in $r$ is present.
The envelope parameter $\sfk$ for both simulated modes is set to $149.12$, a value estimated such that the corresponding envelopes consist of about eight wavelengths along the bent computational domain.

\begin{figure}[htb]
    \centering
    \begin{subfigure}{0.49\textwidth}
        \includegraphics[width=\textwidth]{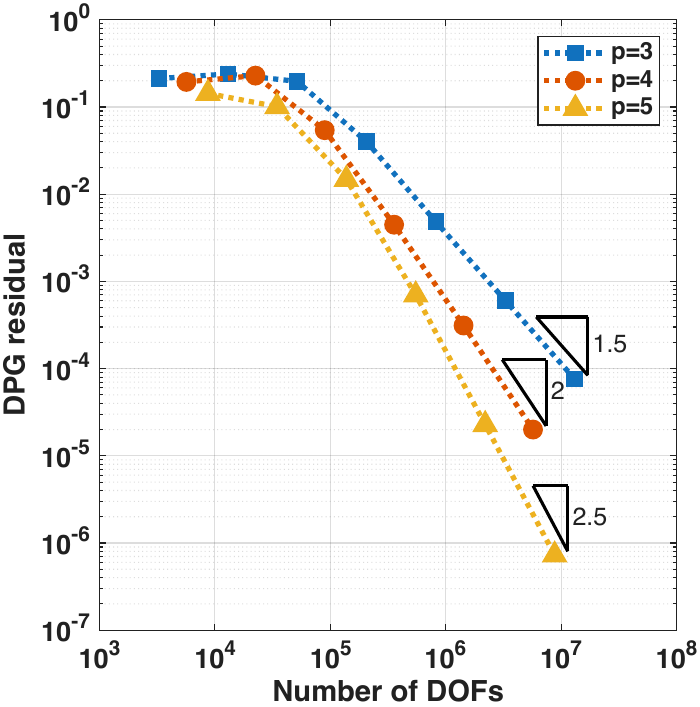}
        \caption{DPG residual}
        \label{plot:ex1-r1300-mode1-residual}
    \end{subfigure}%
    \begin{subfigure}{0.49\textwidth}
        \includegraphics[width=\textwidth]{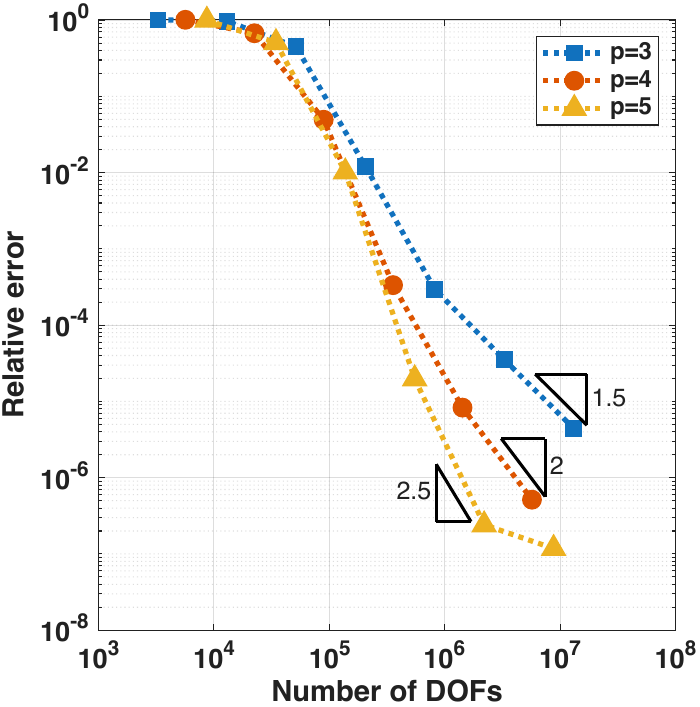}
        \caption{Relative error}
        \label{plot:ex1-r1300-mode1-error}
    \end{subfigure}%
    \caption{Convergence of the DPG residual and relative $L^2$ field error with uniform $h$-refinements and fixed polynomial order, $p \in \{ 3, 4, 5 \}$, for the first mode of the vacuum waveguide with PEC boundary and bend radius $r_0 = 1300$. Note that convergence rates are $h^{p/2}$.}
    \label{fig:ex1-r1300-mode1-conv}
\end{figure}

\begin{figure}[htb]
    \centering
    \begin{subfigure}{0.49\textwidth}
        \includegraphics[width=\textwidth]{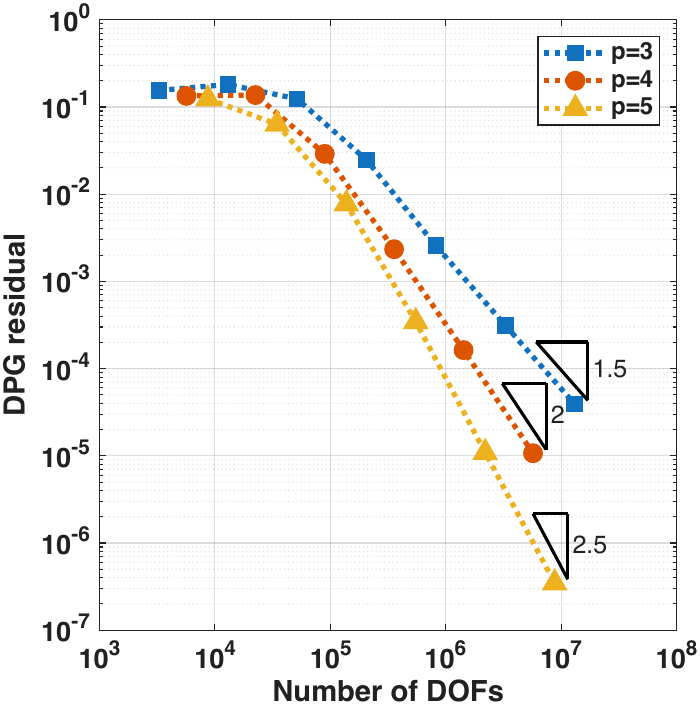}
        \caption{DPG residual}
        \label{plot:ex1-r1300-mode2-residual}
    \end{subfigure}%
    \begin{subfigure}{0.49\textwidth}
        \includegraphics[width=\textwidth]{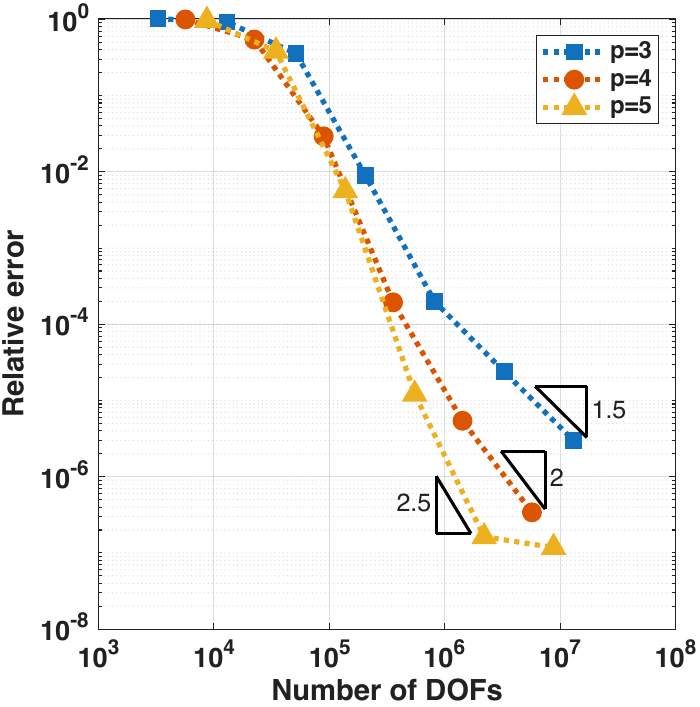}
        \caption{Relative error}
        \label{plot:ex1-r1300-mode2-error}
    \end{subfigure}%
    \caption{Convergence of the DPG residual and relative $L^2$ field error with uniform $h$-refinements and fixed polynomial order, $p \in \{ 3, 4, 5 \}$, for the second mode of the vacuum waveguide with PEC boundary and bend radius $r_0 = 1300$. Note that convergence rates are $h^{p/2}$.}
    \label{fig:ex1-r1300-mode2-conv}
\end{figure}

The solution's approximation order varies within $p\in\{3,4,5\}$; the parameter for the enriched DPG test space is $dp=1$; and the parameter for the DPG test norm is set to $\alpha = 10^{-2}$. 
The initial mesh has 32 exact-geometry hexahedral elements. We apply several uniform anisotropic refinements, in the radial and circumferential coordinates. The convergence of the DPG residual (global norm of $\psi_r$) and the relative field error plotted against the number of degrees of freedom (DOFs), as the mesh is refined, is shown in Figures~\ref{fig:ex1-r1300-mode1-conv} and \ref{fig:ex1-r1300-mode2-conv} for the first mode and the second mode, respectively.

The convergence plots of the DPG residual and relative error demonstrate that asymptotically our implementation attains the expected rates for uniform $h$-refinements. The observed asymptotic rates match $p/2$ (instead of ${p/3}$) because the refinements are performed only in two directions (anisotropically), since the mode profile is constant in the third direction. As a final observation, we note that there is a visible drop in the convergence trend at the last point of the relative error plot for $p=5$ (in plot b of Figures \ref{fig:ex1-r1300-mode1-conv} and \ref{fig:ex1-r1300-mode2-conv}). This dropoff is likely caused by roundoff errors due to the double-precision limit present in our implementation (the field error is an $L^2$-norm that is computed as a square root of a sum of element-local values accumulated in double precision).


\subsection{Numerical experiment 2: Step-index open slab waveguide}
\label{subsec:experiment2}

\begin{figure}[htb]
    \begin{subfigure}{0.48\textwidth}
        \centering
        \includegraphics[width=0.98\linewidth]{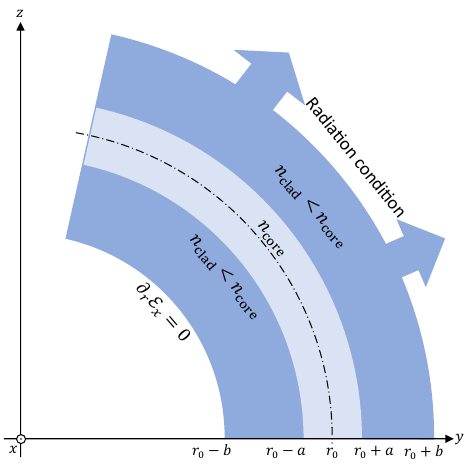}\\[20pt]
        \label{fig:bent_open_slab}
    \end{subfigure}
    \begin{subfigure}{0.52\textwidth}
        \centering
        \includegraphics[width=0.97\linewidth]{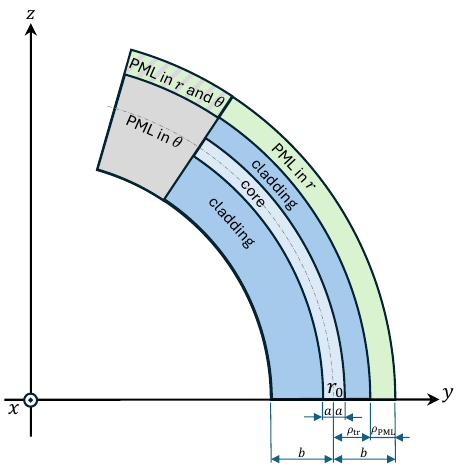}
        \label{fig:ex2-computational-regions}
    \end{subfigure}
    \caption{Setup of numerical experiment 2: (left) diagram of geometry, refractive index distribution, and boundary conditions at the lateral faces; (right) diagram of the distinct computational regions in the simulation. The transverse PML, of width $\rho_{\PML}$, depends on coordinate $r$ only and is applied to a portion of the outer cladding, beginning at $r=r_0+\rho_{\tr}$ and ending at $r=r_0+b$. The parameters that describe the size and location of the PML in $\theta$ are not explicitly marked here as those have already been illustrated in Figure~\ref{fig:pml_theta_diagram}.}
    \label{fig:ex2-setup}
\end{figure}

Next, we compute the envelope Maxwell model solutions for the propagating modes of a bent step-index slab waveguide (also referred to as a three-layer circular waveguide).
The waveguide domain is given by $(-1/2, 1/2) \times (\rbend - b, \rbend + b) \times (0, \theta_{\max})$.
Similar to the first experiment, this problem has an essentially two-dimensional structure, with a constant solution in the third direction.
The waveguide has a core of width $2a$ centered at $r = \rbend$, and a cladding layer of width $b - a$ on each side of the core. 
The refractive index is consequently defined piecewise, $n(r) = \ncore$ for $r \in [\rbend - a, \rbend + a]$, and $n(r) = \nclad$ elsewhere.
The differential equation for the reduced one-dimensional problem still corresponds to \eqref{eq:bessel-problem}. 
The chosen boundary conditions are $\partial \mcE_x/\partial r = 0$ at the left endpoint, while we want to apply a radiation condition at the outer radial boundary, to incorporate the effect of an infinitely wide outer cladding (hence the \emph{open} nature of the waveguide). 
This setup is illustrated in Figure~\ref{fig:ex2-setup} (left).

The radiation condition is imposed via a radial PML region for $r \in \big( \rbend + (a + b) / 2, \rbend + b \big]$, so we can close the system with $\mcE_x=0$ at $r=\rbend+b$. 
Moreover, there are interface conditions in place: $[\![\mcE_x]\!] = 0$ and $[\![d\mcE_x/dr]\!] = 0$ at $r = \rbend \pm a$ ($[\![\cdot]\!]$ is the jump operator). 
A detailed derivation and a comprehensive set of results \revS{for} the semi-analytical computation of the modes of this problem \revS{are} presented in \cite{mora2025bessel}. 
For the present work, we import the eigenvalues and eigenfunctions determined by means of the arbitrary-precision Julia implementation developed therein.\footnote{The mentioned code is available at \url{https://github.com/cgt3/WaveguideSolver}.} 
The mode profiles are furthermore traced along the PML complex path, so that we can enforce the expected decay along the PML direction. 

\begin{figure}[htbp]
    \centering
    \begin{subfigure}{0.49\textwidth}
        \includegraphics[width=\textwidth]{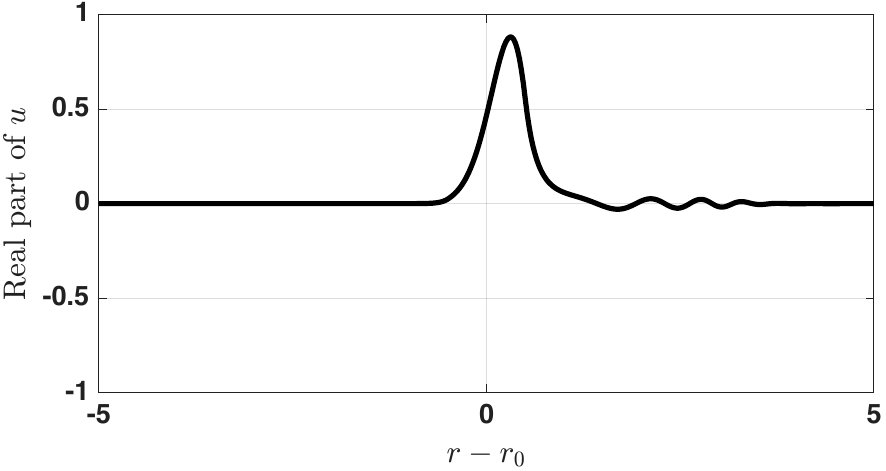}
        \caption{First even mode, $r_0 = 1300$}
        \label{plot:ex1-r1300-even1-real}
    \end{subfigure}%
    \begin{subfigure}{0.49\textwidth}
        \includegraphics[width=\textwidth]{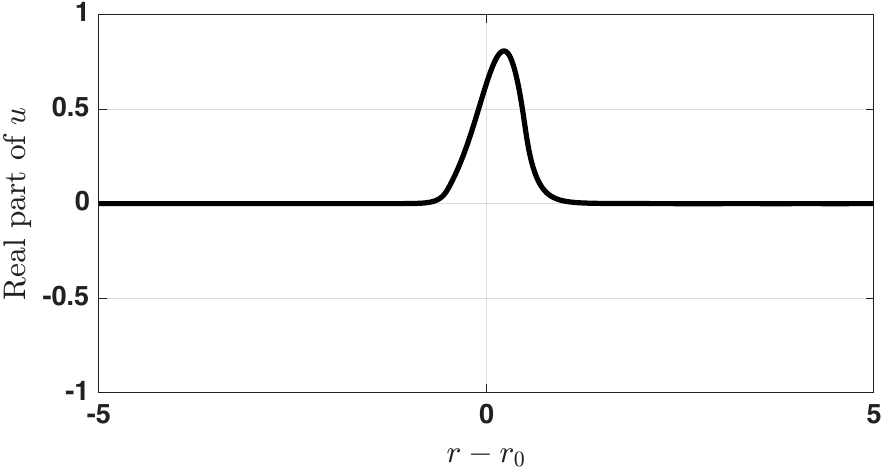}
        \caption{First even mode, $r_0 = 2600$}
        \label{plot:ex1-r2600-even1-real}
    \end{subfigure}
    \begin{subfigure}{0.49\textwidth}
        \includegraphics[width=\textwidth]{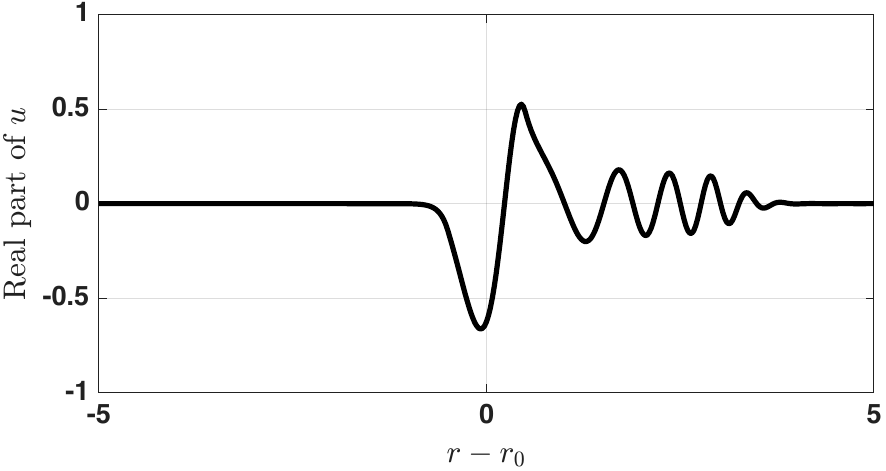}
        \caption{First odd mode, $r_0 = 1300$}
        \label{plot:ex1-r1300-odd1-real}
    \end{subfigure}%
    \begin{subfigure}{0.49\textwidth}
        \includegraphics[width=\textwidth]{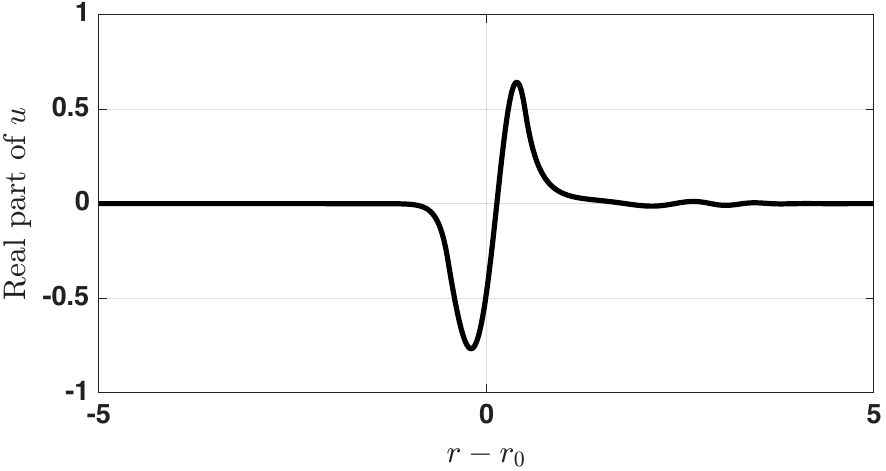}
        \caption{First odd mode, $r_0 = 2600$}
        \label{plot:ex1-r2600-odd1-real}
    \end{subfigure}
    \begin{subfigure}{0.49\textwidth}
        \includegraphics[width=\textwidth]{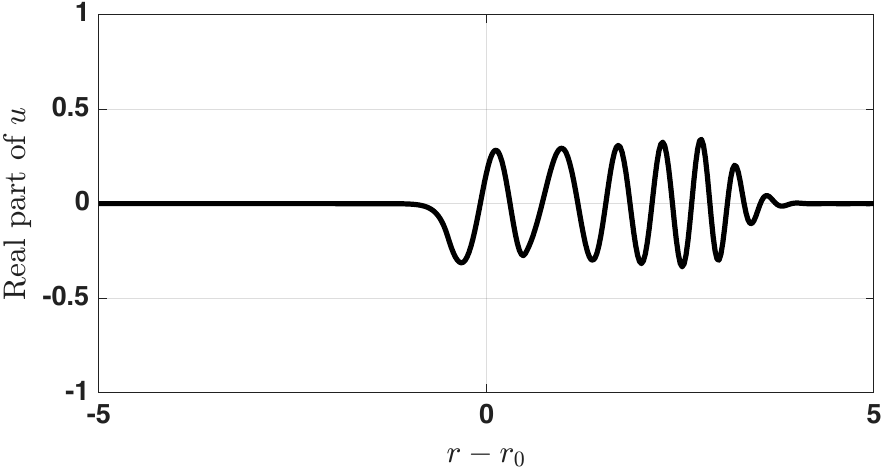}
        \caption{Second even mode, $r_0 = 1300$}
        \label{plot:ex1-r1300-even2-real}
    \end{subfigure}
    \begin{subfigure}{0.49\textwidth}
        \includegraphics[width=\textwidth]{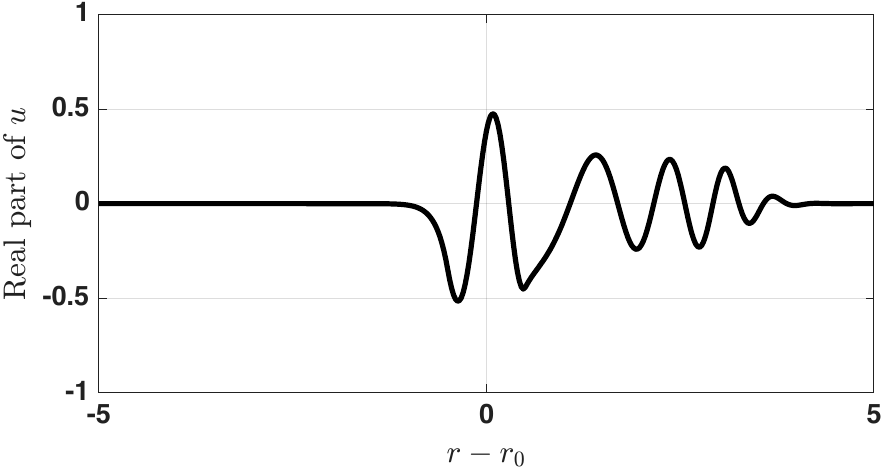}
        \caption{Second even mode, $r_0 = 2600$}
        \label{plot:ex1-r2600-even2-real}
    \end{subfigure}
    \caption{Real components of the mode profiles of the bent step-index slab waveguide, with $r_0 \in \{ 1300, 2600 \}$.} 
    \label{fig:ex2-modes-real}
\end{figure}

\begin{figure}[htbp]
    \centering
    \begin{subfigure}{0.49\textwidth}
        \includegraphics[width=\textwidth]{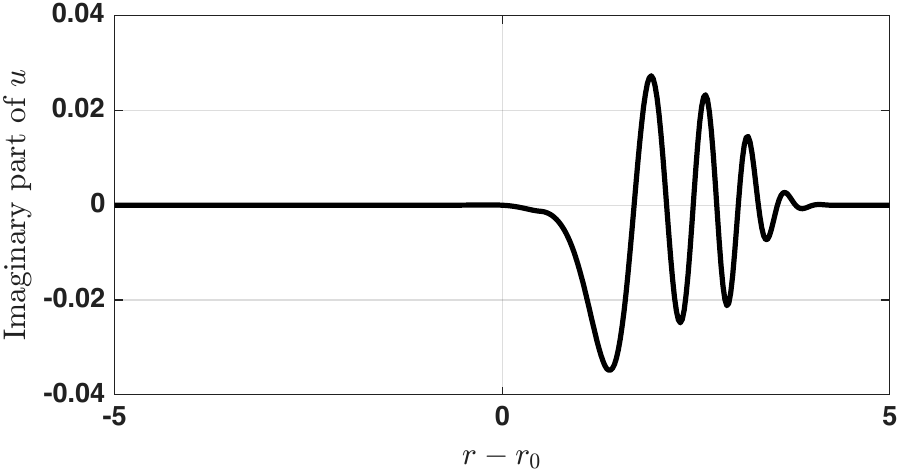}
        \caption{First even mode, $r_0 = 1300$}
        \label{plot:ex1-r1300-even1-imag}
    \end{subfigure}%
    \begin{subfigure}{0.49\textwidth}
        \includegraphics[width=\textwidth]{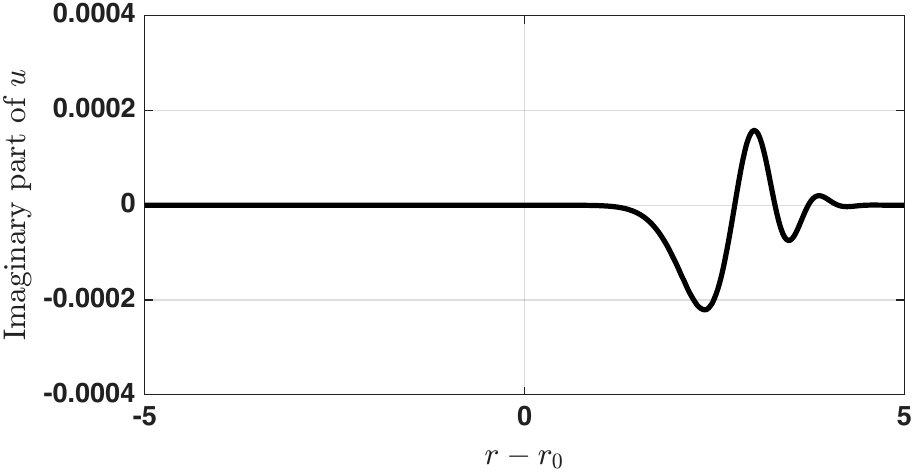}
        \caption{First even mode, $r_0 = 2600$}
        \label{plot:ex1-r2600-even1-imag}
    \end{subfigure}
    \begin{subfigure}{0.49\textwidth}
        \includegraphics[width=\textwidth]{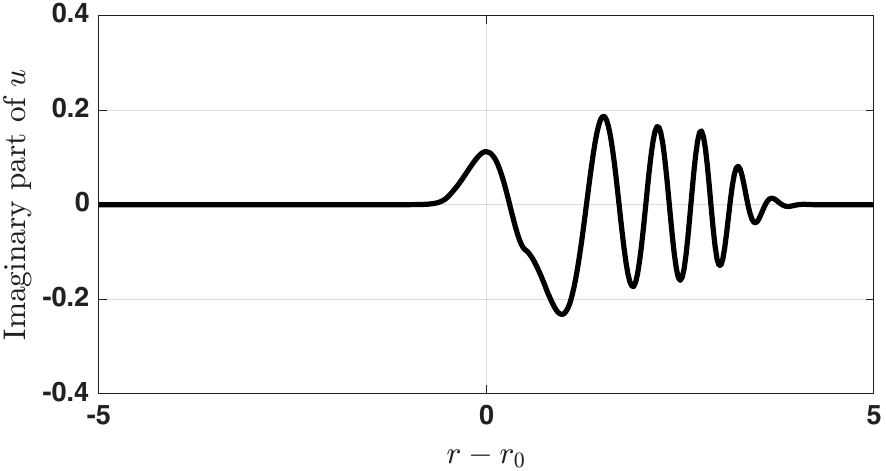}
        \caption{First odd mode, $r_0 = 1300$}
        \label{plot:ex1-r1300-odd1-imag}
    \end{subfigure}%
    \begin{subfigure}{0.49\textwidth}
        \includegraphics[width=\textwidth]{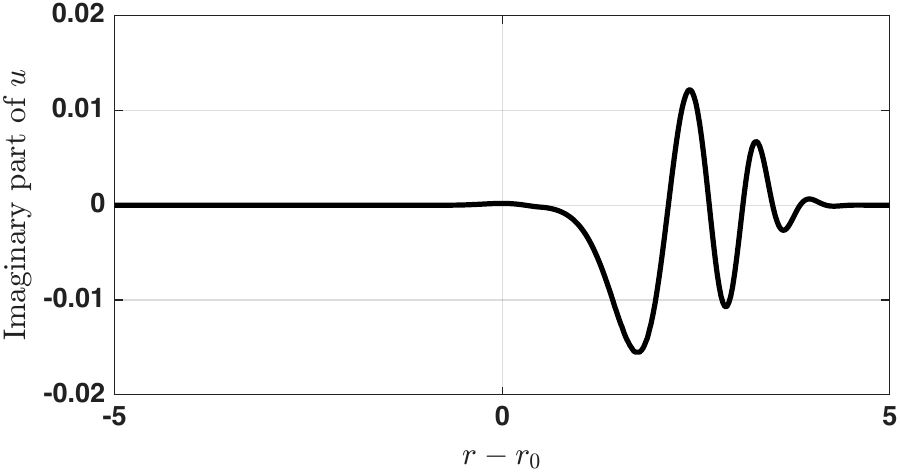}
        \caption{First odd mode, $r_0 = 2600$}
        \label{plot:ex1-r2600-odd1-imag}
    \end{subfigure}
    \begin{subfigure}{0.49\textwidth}
        \includegraphics[width=\textwidth]{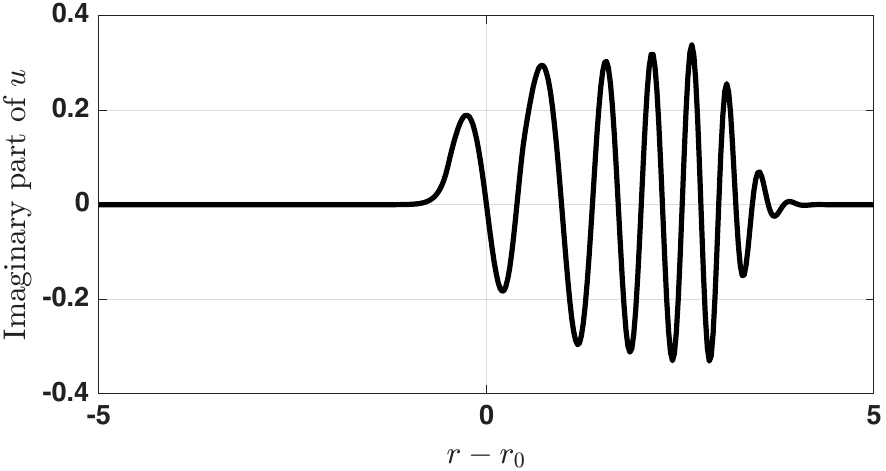}
        \caption{Second even mode, $r_0 = 1300$}
        \label{plot:ex1-r1300-even2-imag}
    \end{subfigure}
    \begin{subfigure}{0.49\textwidth}
        \includegraphics[width=\textwidth]{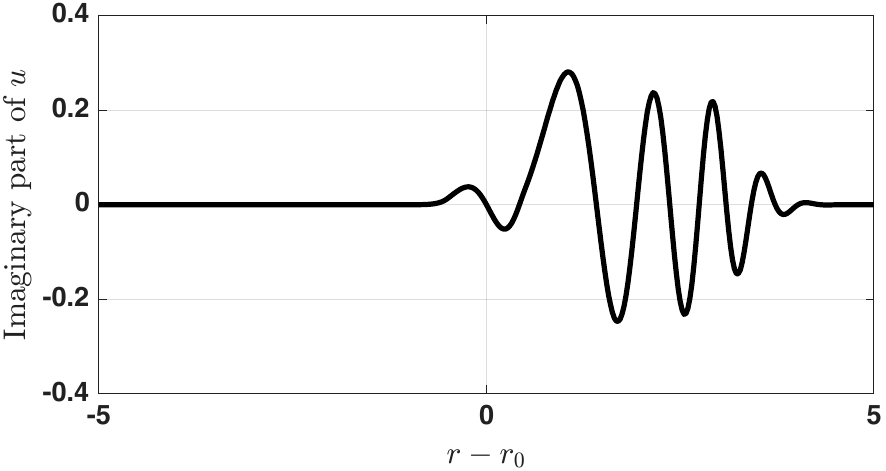}
        \caption{Second even mode, $r_0 = 2600$}
        \label{plot:ex1-r2600-even2-imag}
    \end{subfigure}
    \caption{Imaginary components of the mode profiles of the bent step-index slab waveguide, with $r_0 \in \{ 1300, 2600 \}$.} 
    \label{fig:ex2-modes-imag}
\end{figure}

For the three propagating modes and two different bend radii, Figures~\ref{fig:ex2-modes-real} and \ref{fig:ex2-modes-imag} depict the \revS{profiles} of the real and imaginary parts of the solutions, respectively. 
The scaling of these mode profiles was done by normalizing the function values of each $\mcE_x(r)$ with respect to its weighted $L^2$-norm
$$
\left(\int_{r_0-b}^{r_0+b}\frac{r_0}{r} | \mcE_x(r)|^2 dr\right)^{1/2}.
$$

In our finite element simulations, the boundary condition at $r = \rbend - b$ is implemented indirectly, by imposing the homogeneous condition on the magnetic field trace, $\bfn\times\check{\sfH} = 0$.
Besides this \emph{magnetic trace boundary condition} and the radial PML discussed above, the remaining boundary conditions are implemented \revS{analogously to those in} the first numerical example\revisedB{; that is, $\bfn \times \check{\sfE} = 0$ on the front and back faces; similarly, $\bfn \times \check{\sfE} = 0$ on the top and outer radial boundary past the PMLs, and $\bfn \times \check{\sfE} = \bfn \times (\mcE_x(r) \bfe_x)$ on the bottom face}.
The interface conditions need not be implemented explicitly in the DPG code, as the variational formulation enforces the weak satisfaction of Maxwell's equations even if the refractive index is discontinuous. 
A diagram illustrating the geometry and the computational regions for this second numerical experiment is shown in Figure~\ref{fig:ex2-setup} (right).

We perform the simulations for all six cases (three modes by two bend radii) using an initial mesh of 384 hexahedral elements followed by up to 25 adaptive refinements. We use isotropic $p=5$ uniformly, and keep the parameters $dp=1$ and $\alpha=10^{-2}$ the same as in the first experiment. The envelope parameter is set to $\sfk=217.43$, slightly below $k_0\nclad$. Finally, the maximum angle is $\theta_{\max}=9.76^\circ$ and $\theta_{\max}=4.78^\circ$, for $\rbend=1300$ and $\rbend=2600$, respectively.

\subsubsection{Convergence results}

\begin{figure}[htb]
    \centering
    \begin{subfigure}{0.49\textwidth}
        \includegraphics[width=\textwidth]{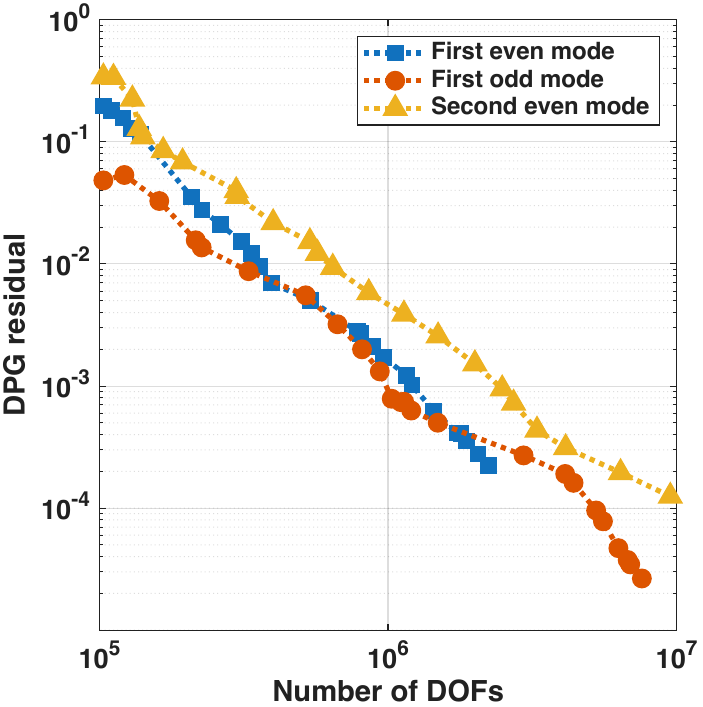}
        \caption{$r_0 = 1300$}
        \label{plot:ex2-r1300-resid-conv}
    \end{subfigure}%
    \begin{subfigure}{0.49\textwidth}
        \includegraphics[width=\textwidth]{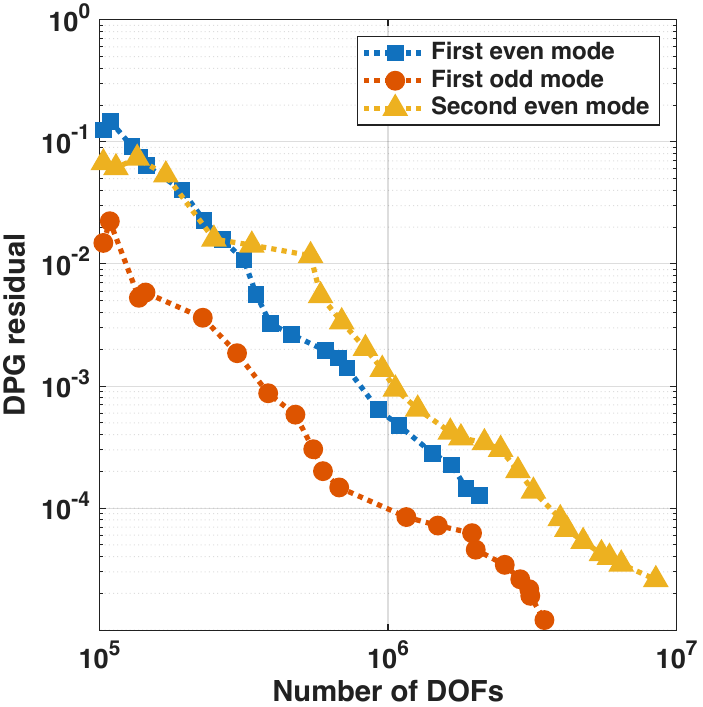}
        \caption{$r_0 = 2600$}
        \label{plot:ex2-r2600-resid-conv}
    \end{subfigure}%
    \caption{Convergence of the DPG residual with adaptive $h$-refinements and fixed polynomial order, $p = 5$, for the three propagating modes of the bent step-index slab waveguide, with $r_0 \in \{ 1300, 2600 \}$.}
    \label{fig:ex2-resid-conv}
\end{figure}

In the first numerical experiment, we showed convergence rates of the numerical error and the DPG residual, where it was noticeable that the decreasing behavior was very similar in both; here, we present the convergence of the residual only. The computation of the error is omitted in this experiment because of the high computational cost of evaluating the quadruple-precision functions (the modes) in the entire domain for each mesh during the adaptive refinements.
Figure~\ref{plot:ex2-r1300-resid-conv} shows the DPG residual plotted against the number of DOFs for the case of bend radius $\rbend=1300$; each line represents one of the first three propagating modes. Figure~\ref{plot:ex2-r2600-resid-conv} shows the corresponding results for $\rbend=2600$. 
We note that in all six cases, the residual decreases by at least three orders of magnitude.

Since a key motivation of our work is the prediction of bend losses, the following subsection demonstrates the excellent alignment of our numerically computed losses with the losses predicted by the semi-analytical solutions derived using the method in \cite{mora2025bessel}.

\subsubsection{Loss factor estimation}

Recall that the Poynting vector $S$ is given by
\begin{equation*}
    S(x, r, \theta) 
    = E(x,r,\theta)\times\overline{H(x,r,\theta)}
    = e^{-\zI \sfk \rbend \theta} \sfE(x, r, \theta) \times \overline{\big( e^{-\zI \sfk \rbend \theta} \sfH(x, r, \theta) \big)}       
    = \sfE(x, r, \theta) \times \overline{\sfH(x, r, \theta)} ,
\end{equation*}
which implies that $S$ is invariant to the envelope wavenumber ansatz \eqref{eq:envelope-ansatz}. Then, introducing the TE ansatz
\begin{align*}
    S(x,r,\theta) & = e^{-\zI \beta \theta} \mcE(x, r) \times \overline{\big( e^{-\zI\beta\theta} \mcH(x, r) \big)} \\
        & = e^{-\zI \beta \theta + \zI \overline{\beta} \theta} \mcE(x, r) \times \overline{\mcH(x, r)} \\
        & = e^{2 \Im(\beta) \theta} \mcE(x, r) \times \overline{\mcH(x, r)} ,
\end{align*}
we note that a negative imaginary part of the propagation constant leads to an exponential decay of the magnitude of $S$ for increasing $\theta$. 
This exponential coefficient for the decay of $S$ is twice as large as that of $E$ or $H$. We interpret $\Im(\beta)$ as the electromagnetic field's amplitude attenuation or loss per unit radian. Now, recall the definition of the irradiance:
\begin{align}
    I(x,r,\theta) 
    &= \big| \Re\big( S(x, r, \theta) \big) \big|  \label{eq:te-irradiance1} \\
    &= e^{2 \Im(\beta) \theta} \big| \Re\big( \mcE(x, r) \times \overline{\mcH(x, r)} \big) \big| . \label{eq:te-irradiance2}
\end{align}
Integrating $I$ over the cross-section $A$ at a given angle $\theta$, we define the optical power by
\begin{equation}
    \mcP(\theta)=\int_A I(x,r,\theta) dxdr.
    \label{te-optical-power}
\end{equation}
Note that the ratio of $\mcP(\theta)$ and $\mcP(0)$ satisfies
$$
e^{2\Im(\beta)\theta}=\frac{\mcP(\theta)}{\mcP(0)}.
$$
The numerical approximation of $\mcP(\theta)$ is directly computable from the finite element trace solutions $\check{\sfE}_h,\check{\sfH}_h$, through
\begin{equation}
    \mcP_h(\theta)=\int_A |\Re(\check{\sfE}_h(x,r,\theta)\times\overline{\check{\sfH}_h(x,r,\theta)}) |d xdr.
    \label{eq:discrete-optical-power}
\end{equation}
Using \eqref{eq:te-irradiance2}--\eqref{eq:discrete-optical-power}, and given any $\theta$, we can estimate the
\begin{equation}
    \text{numerical loss factor} := 
    \frac{1}{2\theta}\ln\left(\frac{\mcP_h(\theta)}{\mcP_h(0)}\right)
    \approx
    \frac{1}{2\theta}\ln\left(\frac{\mcP(\theta)}{\mcP(0)}\right)
    =
    \Im(\beta).
    \label{eq:num-loss-factor}
\end{equation}

Using \eqref{eq:num-loss-factor}, we can estimate numerically, from the 3D envelope Maxwell model solutions, the imaginary component of the propagation constant. 
The \revS{resulting} estimates are then compared \revS{with} the loss factors computed \revS{using} the high-precision eigensolver of \cite{mora2025bessel}. 
With this approach, we can create loss factor convergence plots for different modes and bend radii, as the meshes are adaptively refined.

\begin{figure}[htbp]
    \centering
    \begin{subfigure}{0.43\textwidth}
        \includegraphics[width=\textwidth]{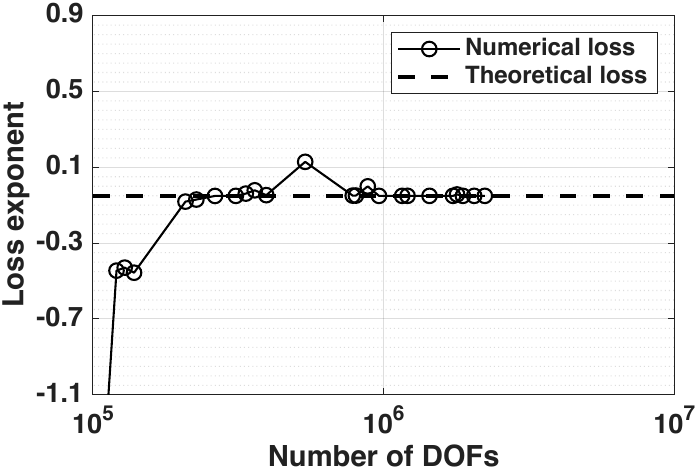}
        \caption{\centering{First even mode, $r_0 = 1300$.\\ Theoretical loss exponent = $-5.0712\cdot10^{-2}$.\\
        Final numerical loss exponent = $-5.0628\cdot10^{-2}$.}}
        \label{plot:ex2-r1300-loss-exp-even1}
    \end{subfigure}%
    \hspace{0.02\textwidth}%
    \begin{subfigure}{0.43\textwidth}
        \includegraphics[width=\textwidth]{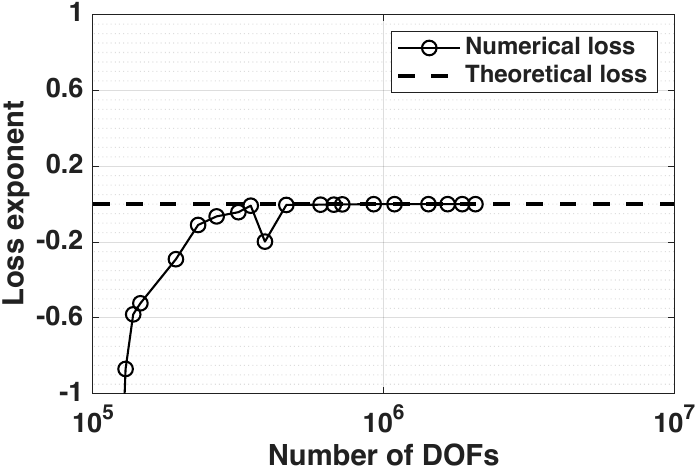}
        \caption{\centering{First even mode, $r_0 = 2600$.\\ Theoretical loss exponent = $-3.2118\cdot10^{-6}$.\\
        Final numerical loss exponent = $-1.2175\cdot10^{-4}$.}}
        \label{plot:ex2-r2600-loss-exp-even1}
    \end{subfigure}
    \begin{subfigure}{0.43\textwidth}
        \includegraphics[width=\textwidth]{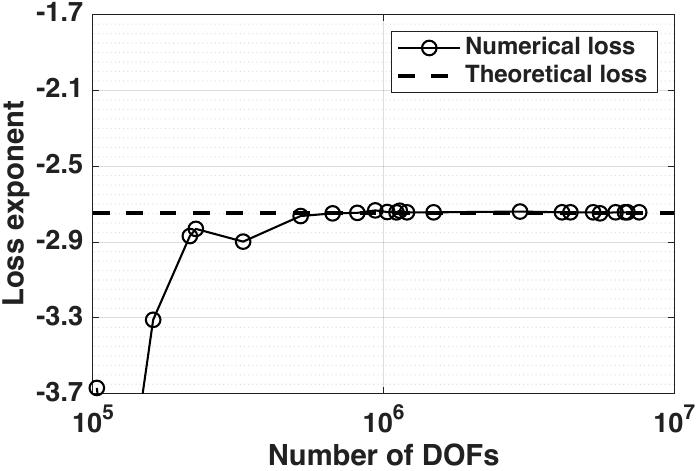}
        \caption{\centering{First odd mode, $r_0 = 1300$.\\ Theoretical loss exponent = $-2.74478$.\\
        Final numerical loss exponent = $-2.74303$.}}
        \label{plot:ex2-r1300-loss-exp-odd1}
    \end{subfigure}%
    \hspace{0.02\textwidth}%
    \begin{subfigure}{0.43\textwidth}
        \includegraphics[width=\textwidth]{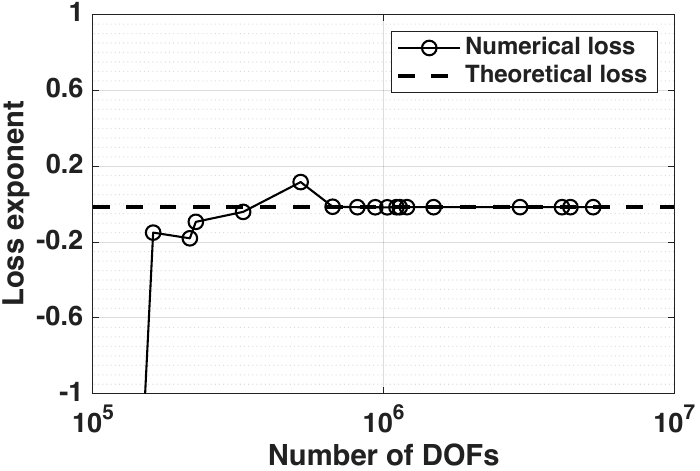}
        \caption{\centering{First odd mode, $r_0 = 2600$.\\ Theoretical loss exponent = $-1.5924\cdot10^{-2}$.\\
        Final numerical loss exponent = $-1.5878\cdot10^{-2}$.}}
        \label{plot:ex2-r2600-loss-exp-odd1}
    \end{subfigure}
    \begin{subfigure}{0.43\textwidth}
        \includegraphics[width=\textwidth]{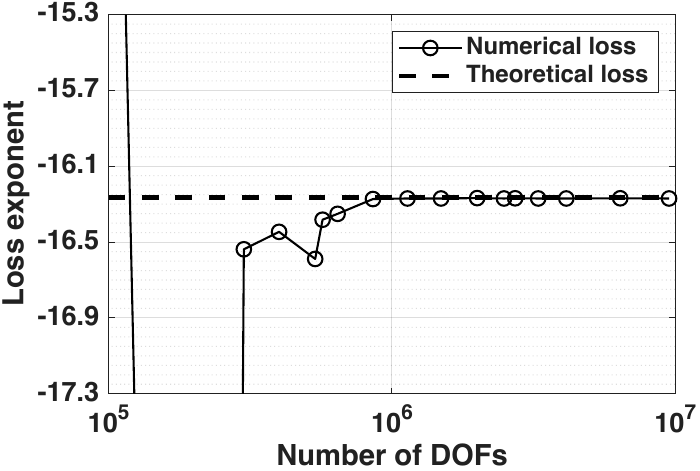}
        \caption{\centering{Second even mode, $r_0 = 1300$.\\
        Theoretical loss exponent = $-16.2649$.\\
        Final numerical loss exponent = $-16.2698$.}}
        \label{plot:ex2-r1300-loss-exp-even2}
    \end{subfigure}%
    \hspace{0.02\textwidth}%
    \begin{subfigure}{0.43\textwidth}
        \includegraphics[width=\textwidth]{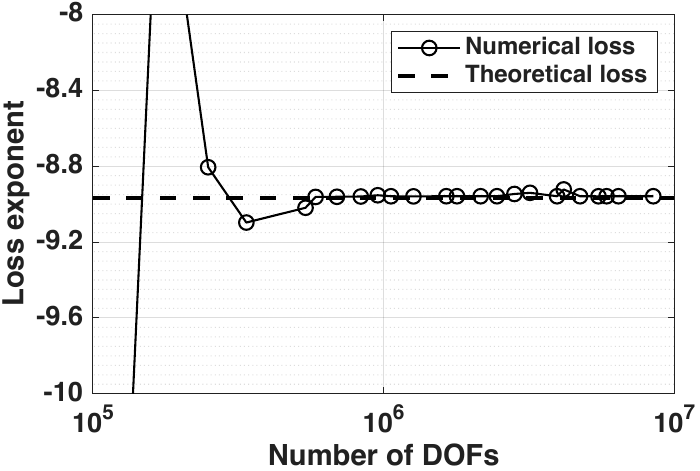}
        \caption{\centering{Second even mode, $r_0 = 2600$.\\
        Theoretical loss exponent = $-8.9680$.\\
        Final numerical loss exponent = $-8.9582$.}}
        \label{plot:ex2-r2600-loss-exp-even2}
    \end{subfigure}
    \caption{Numerical computation of loss factor for the three propagating modes of the bent step-index slab waveguide, with $r_0 \in \{ 1300, 2600 \}$. All plots show how the numerical loss values converge to the theoretical estimates. In the caption of each of the six plots, we present the final numerical loss exponent next to the theoretical loss exponent, demonstrating that only in case (b) there is a major relative discrepancy.}
    \label{fig:ex2-loss-exp}
\end{figure}

This comparison of the loss exponent is presented in Figure~\ref{fig:ex2-loss-exp}. In all cases, formula \eqref{eq:num-loss-factor} is evaluated at the PML transition point, $\theta=\theta_{\tr}$; and the base value of $\mcP_h(0)$ (input facet optical power) is computed on the finest mesh in each scenario.
In the figure, plots \subref{plot:ex2-r1300-loss-exp-even1}, \subref{plot:ex2-r1300-loss-exp-odd1} and \subref{plot:ex2-r1300-loss-exp-even2} show the results for all three modes for $\rbend=1300$, while plots \subref{plot:ex2-r2600-loss-exp-even1}, \subref{plot:ex2-r2600-loss-exp-odd1} and \subref{plot:ex2-r2600-loss-exp-even2} do so for $\rbend=2600$. The caption of each plot specifies the semi-analytically derived value of the corresponding loss exponent. 

In all cases, the numerically estimated loss factors appear to converge to the theoretical/semi-analytic values in terms of absolute error.
We note that for the first even mode, the theoretical loss exponent is extremely small, leading to larger \emph{relative} errors in the numerically estimated values. 
For instance, with $\rbend=2600$, the numerical loss estimates are on the order of $\mcO(10^{-4})$, whereas the semi-analytical value is $\Im(\beta)=-3.2118\cdot 10^{-6}$ (requiring $\sim10^6$ radians to observe significant attenuation in this mode's amplitude). 
Bend losses become a significant factor \revS{in} mode propagation only for considerably larger values of the loss exponent (e.g., $\Im(\beta)=\mcO(1)$ or larger). 
In these cases, the numerical experiments (first odd mode with $\rbend=1300$, second even mode with both bend radii) indicate excellent agreement of the numerically estimated loss factors with the semi-analytic values.


\subsection{Numerical experiment 3: Bent step-index optical fiber}
\label{subsec:experiment3}

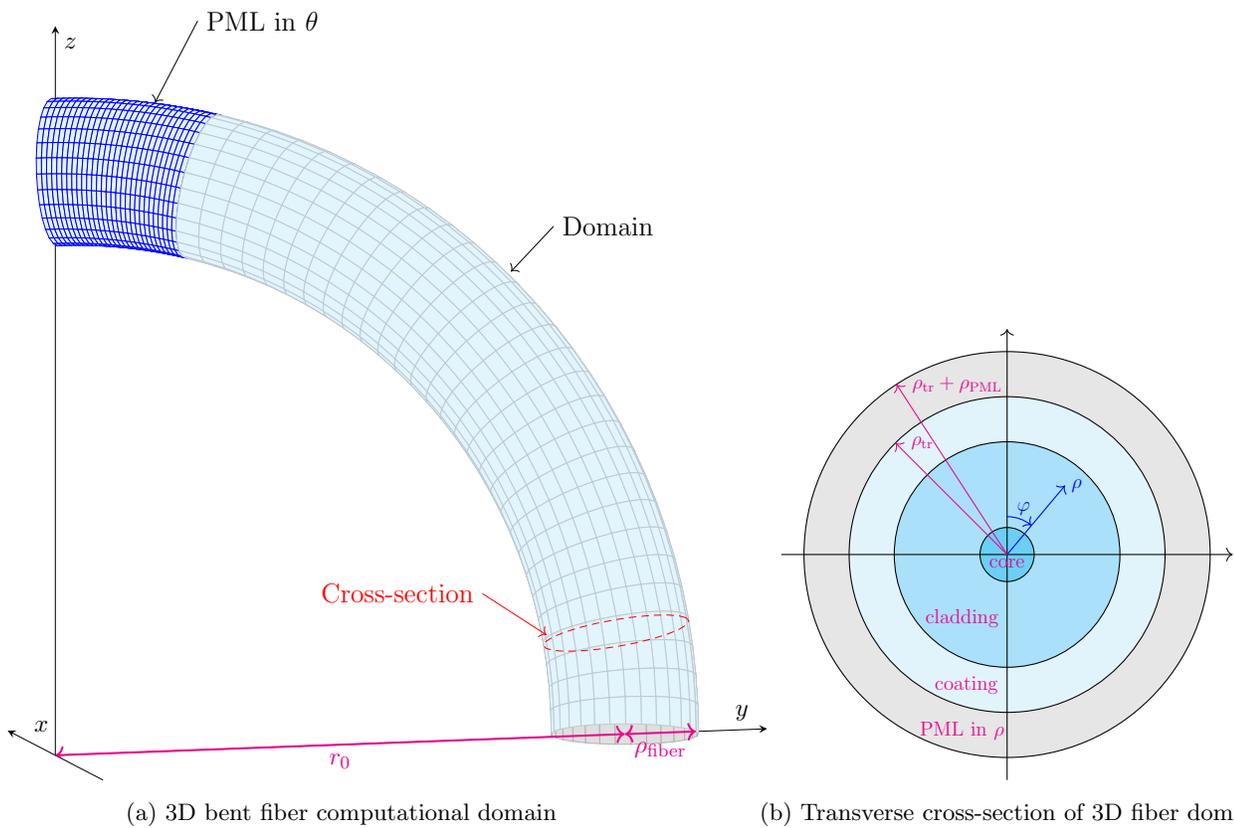
\begin{figure}[htbp]
    \centering
    \begin{subfigure}{0.5\textwidth}        
    \pgfplotsset{width=16cm,compat=1.18}
    \begin{tikzpicture}[scale=0.95, every node/.style={scale=0.9}]
        \def\thup{72}
        \def\pmlth{15}
        \def\rtor{5}
        \def\rbnd{40}
        \begin{axis}[
          view={105}{-8},
          axis equal image,
          axis lines=middle,
          xmin=-2.5*\rtor,ymin=0,zmin=0,
          xmax=2.5*\rtor,ymax=\rbnd+2*\rtor,zmax=\rbnd+2*\rtor,
          ticks=none,
          mesh/interior colormap={gy}{color=(gray!25) color=(gray!25)},
          colormap={bl}{color=(cyan!10) color=(cyan!10)},
          xlabel=$x$, ylabel=$y$, zlabel=$z$
        ]
        \addplot3[domain=0:360,y domain=75:90, samples=30,
        surf,shader=faceted,faceted color=blue,z buffer=sort]
        ({\rtor * sin(x)},
        {(\rbnd + \rtor * cos(x)) * cos(y)} ,
        {(\rbnd + \rtor * cos(x)) * sin(y)});
        \addplot3[domain=0:359,y domain=0:75, samples=30,
        surf,shader=faceted,z buffer=sort]
        ({\rtor * sin(x)},
        {(\rbnd + \rtor * cos(x)) * cos(y)} ,
        {(\rbnd + \rtor * cos(x)) * sin(y)});
    

        \draw [<->,thick,magenta] (axis cs: 0,0,0) -- (axis cs: 0,\rbnd,0) node [midway,below] {$\rbend$};
        \draw [<->,thick,magenta] (axis cs: 0,\rbnd,0) -- (axis cs: 0,\rbnd+\rtor,0) node [midway,below] {$\rfi$};
        \draw [->] (axis cs: 0,10,50) -- node[pos=0.0,right] {\large PML in $\theta$} (axis cs: 0,7,44.5) ;
        \draw [->] (axis cs: 0,35,35) -- node[pos=0.0,right] {\large Domain} (axis cs: 0,32,32) ;
        \draw [->,red] (axis cs: 0,30,10) -- node[pos=0.0,left] {\large Cross-section} (axis cs: 0,34.5,7) ;
        \addplot3 [red,densely dashed,domain=0:360,samples y=1] 
        ({\rtor * sin(x)},
        {(\rbnd + \rtor * cos(x)) * cos(10)} ,
        {(\rbnd + \rtor * cos(x)) * sin(10)});
      \end{axis}
    \end{tikzpicture}
    \caption{3D bent fiber computational domain}
    \label{fig:exp3-domain}
    \end{subfigure}
    \begin{subfigure}{0.49\textwidth}
    \centering
    \begin{tikzpicture}[scale=0.6, every node/.style={scale=0.64}]
        
        \def\rcoreval{0.6}
        \def\rcladval{2.5}
        \def\rcoatval{3.5}
        \def\rpmlval{4.5}
        
        \draw[fill = gray!20] (0,0) circle(\rpmlval);
        \draw[fill = cyan!10] (0,0) circle(\rcoatval);
        \draw[fill = cyan!30] (0,0) circle(\rcladval);
        \draw[fill = cyan!50] (0,0) circle(\rcoreval);
        \draw[->, black] (-\rpmlval - 0.5, 0) -- (\rpmlval + 0.5, 0) node[anchor = west] {};
        \draw[->, black] (0, -\rpmlval - 0.5) -- (0, \rpmlval + 0.5) node[anchor = south] {};
    
        \draw[->, blue] (0,0) -- ({0.8 * \rcladval * cos(50)}, {0.8 * \rcladval * sin(50)}) node[anchor = west] {\large $\rho$};
        \draw[->, blue] (0,1.4 * \rcoreval) arc[start angle=90, end angle = 50, radius = 1.4 * \rcoreval];
        \node[blue] at ({1.8 * \rcoreval * cos(70)}, {1.8 * \rcoreval * sin(70)}) {\large $\varphi$};
    
        \node[magenta] at (0.0, -0.2) {\large core};
        \node[magenta] at (-1.0, -1.4){\large cladding};
        \node[magenta] at (-0.9, -2.9) {\large coating};
        \node[magenta] at (-1.0, -3.9) {\large PML in $\rho$};
        \draw[->, magenta] (0,0) -- ({\rcoatval * cos(135)}, {\rcoatval * sin(135)}) node[anchor = west,xshift = 0.5em] {\large $\rho_{\text{tr}}$};
        \draw[->, magenta] (0, 0) -- ({\rpmlval * cos(123)}, {\rpmlval * sin(123)}) node[anchor = west, xshift = 0.5em] {\large $\rho_{\text{tr}} + \rho_{\text{PML}}$};
    \end{tikzpicture}
    \caption{Transverse cross-section of 3D fiber domain}
    \label{fig:exp3-cross-section}
    \end{subfigure}
    \caption{Geometry setup for numerical experiment 3: (a) schematic of the three-dimensional computational domain of the toroidal body that represents a coiled fiber, including the PML in $\theta$; and (b) schematic of the transverse cross-section of the 3D fiber domain, including the PML in $\rho$.}
    \label{fig:exp3-setup}
\end{figure}

In our third numerical experiment, we simulate the propagation of high-order modes through a weakly-guiding bent step-index fiber. 
The geometry setup of the computational experiment is portrayed in Figure~\ref{fig:exp3-setup}. 
The three-dimensional fiber domain, with its toroidal shape, includes a PML in $\theta$ (\ref{fig:exp3-domain}), and the transverse cross-section of the fiber features a PML in $\rho$ (\ref{fig:exp3-cross-section}).
Except for the coating radius, the physical dimensions of the fiber domain and other fiber parameters match the parameters listed in Table~\ref{tab:fiber_data}.
The coating radius in this experiment is decreased to $\tfrac{4}{3}\rclad$. Using this smaller outer diameter reduces the total cross-sectional area, thereby reducing the total computational cost of the fiber simulation. The PML in $\rho$ starts at $\rho_{\tr}=\tfrac{7}{6}\rclad$, leaving a radial PML size of $\rho_{\PML}=\tfrac{1}{6}\rclad$ (that is, half the coating thickness). Placing the PML at this location is a choice based on the expectation that the optical field strongly decays across the fiber coating.

\begin{figure}[htb]
    \centering
    \includegraphics[width=0.4\linewidth]{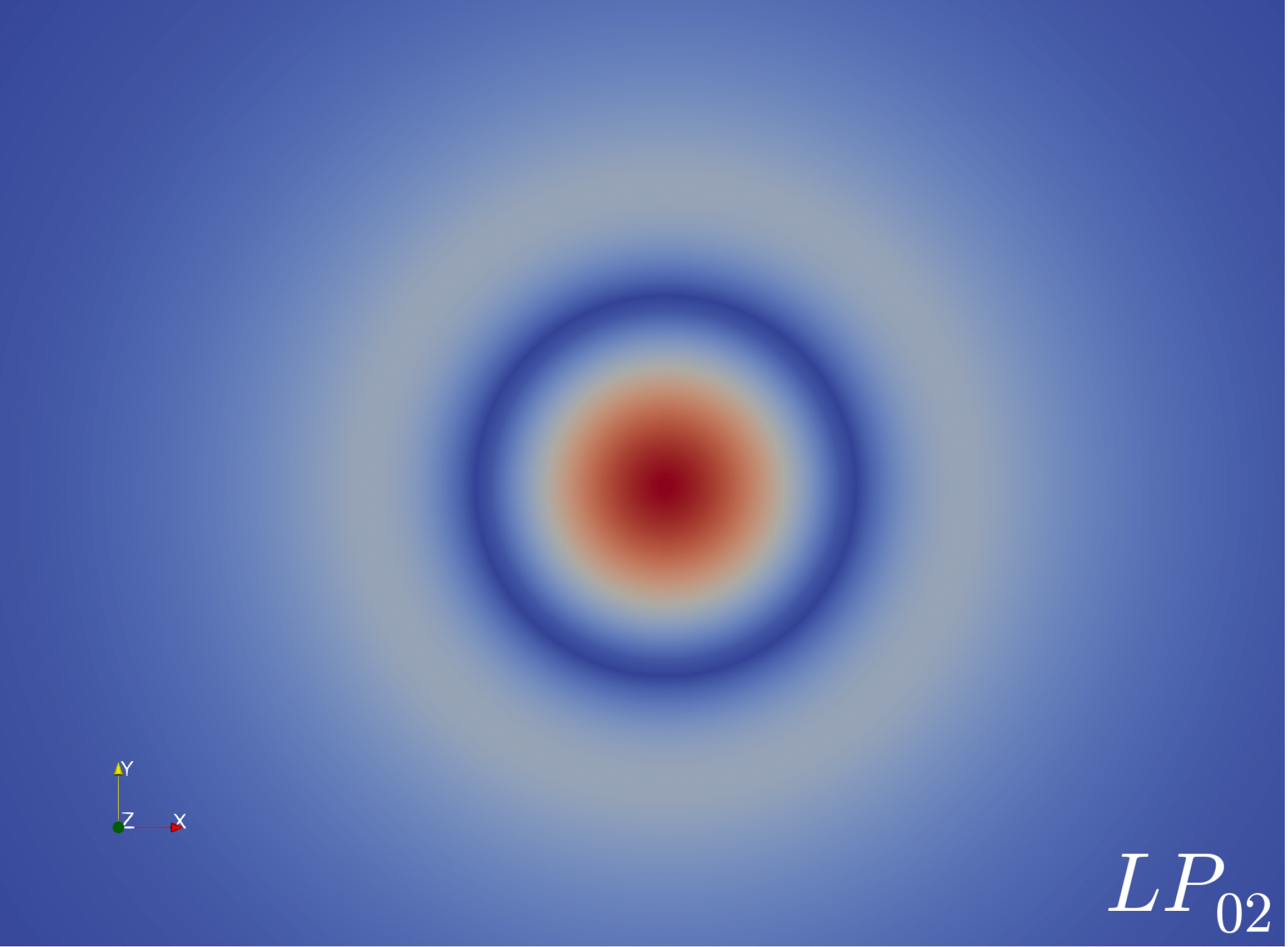}
    \caption{Contour plot of the irradiance of the LP$_{02}$ mode of a straight weakly-guiding step-index optical fiber, with fiber core and cladding parameters listed \revS{in} Table~\ref{tab:fiber_data}. This irradiance plot across the transverse fiber cross-section is a close-up of the core region with a small part of the inner cladding visible.}
    \label{fig:straight-lp02}
\end{figure}

In this experiment, we know neither the exact solution nor the theoretical bend loss. Rather, the problem setup consists of \revisedB{(1)} exciting the waveguide with one of the \revisedB{radially symmetric} modes of the straight fiber\revisedB{, $\mcE_x(\rho)$}, imposed as a boundary condition on the input facet (bottom face, $\theta=0$)\revisedB{: $\nml\times\check{\sfE}=\nml\times(\mcE_x(\rho) \bfe_x)$}; \revisedB{(2)} adding PMLs in $\theta$ and in $\rho$\revisedB{, with their corresponding homogeneous tangential b.c., $\nml\times\check{\sfE}=0$ at $\rho=\rfi$ and at $\theta=\theta_{\max}$}; and \revisedB{(3)} letting the adaptive DPG solutions develop over several mesh \revS{refinement} iterations. As an input mode, we use the radially-symmetric LP$_{02}$ mode of the straight fiber, with the transverse profile illustrated in Figure~\ref{fig:straight-lp02}.

We compute the simulation for two bend radii, $\rbend\in\{1300,2600\}$, fixing the total arc length of the simulated fiber at $\rbend\theta_{\max}=192.73$ non-dimensional length units. 
The numerical results are obtained with DPG parameters $p=4$, $dp=1$ and $\alpha=10$, using envelope wavenumber $\sfk=217.375$. 
The adaptive meshing in this experiment is mainly anisotropic, as we focus more on refining the elements in the transverse directions rather than in the longitudinal direction.

The initial mesh has 8\,192 elements (256 in the cross section $\times$ 32 in the longitudinal direction), combining exact-geometry prisms and hexahedra, and we perform 15 refinements. The final mesh has 46\,403 elements, and a total of 24\,425\,352 degrees of freedom.

\begin{figure}[htbp]
    \centering
    \begin{minipage}{0.5\textwidth}
    \includegraphics[width=\linewidth]{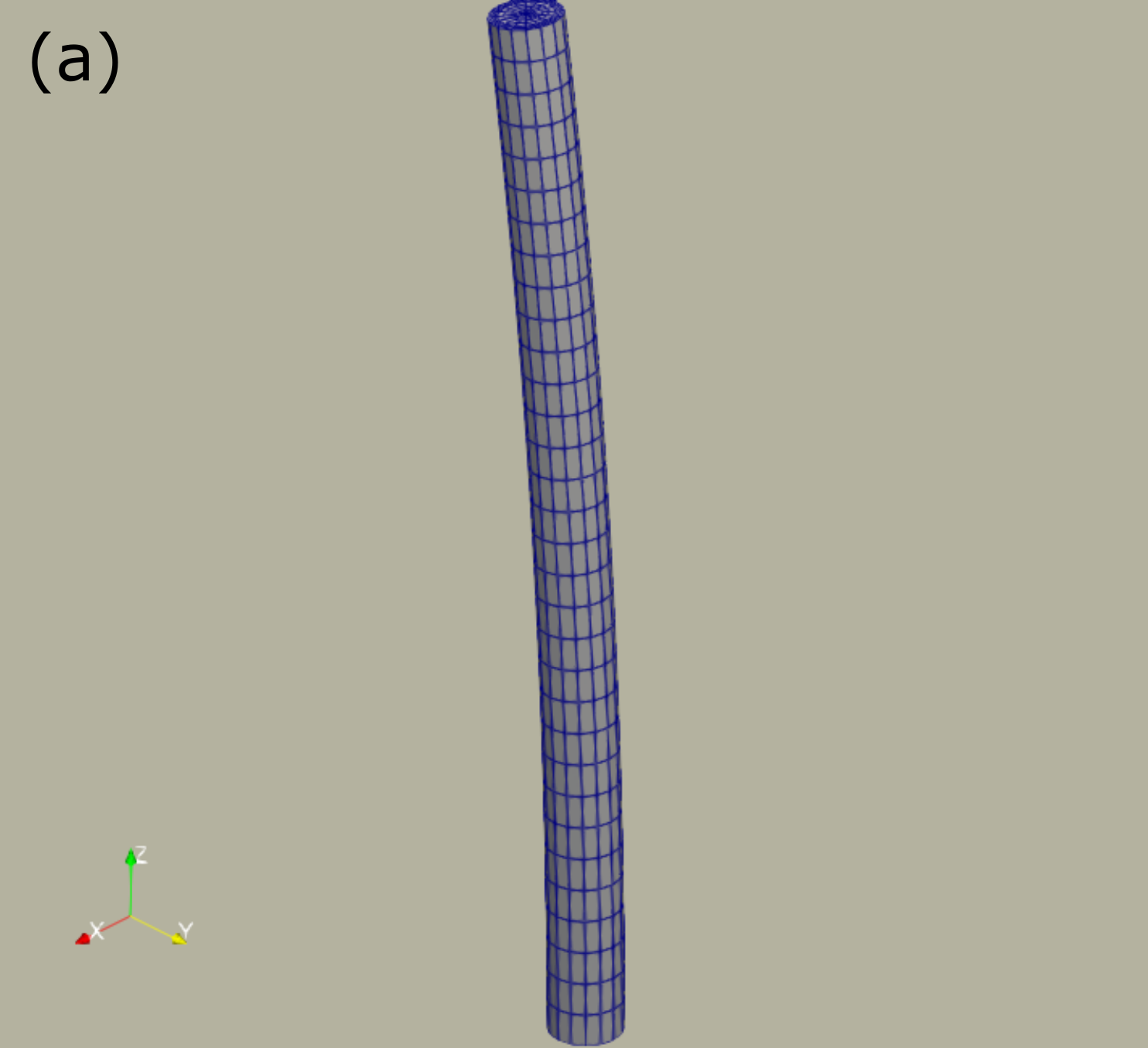}
    \includegraphics[width=\linewidth]{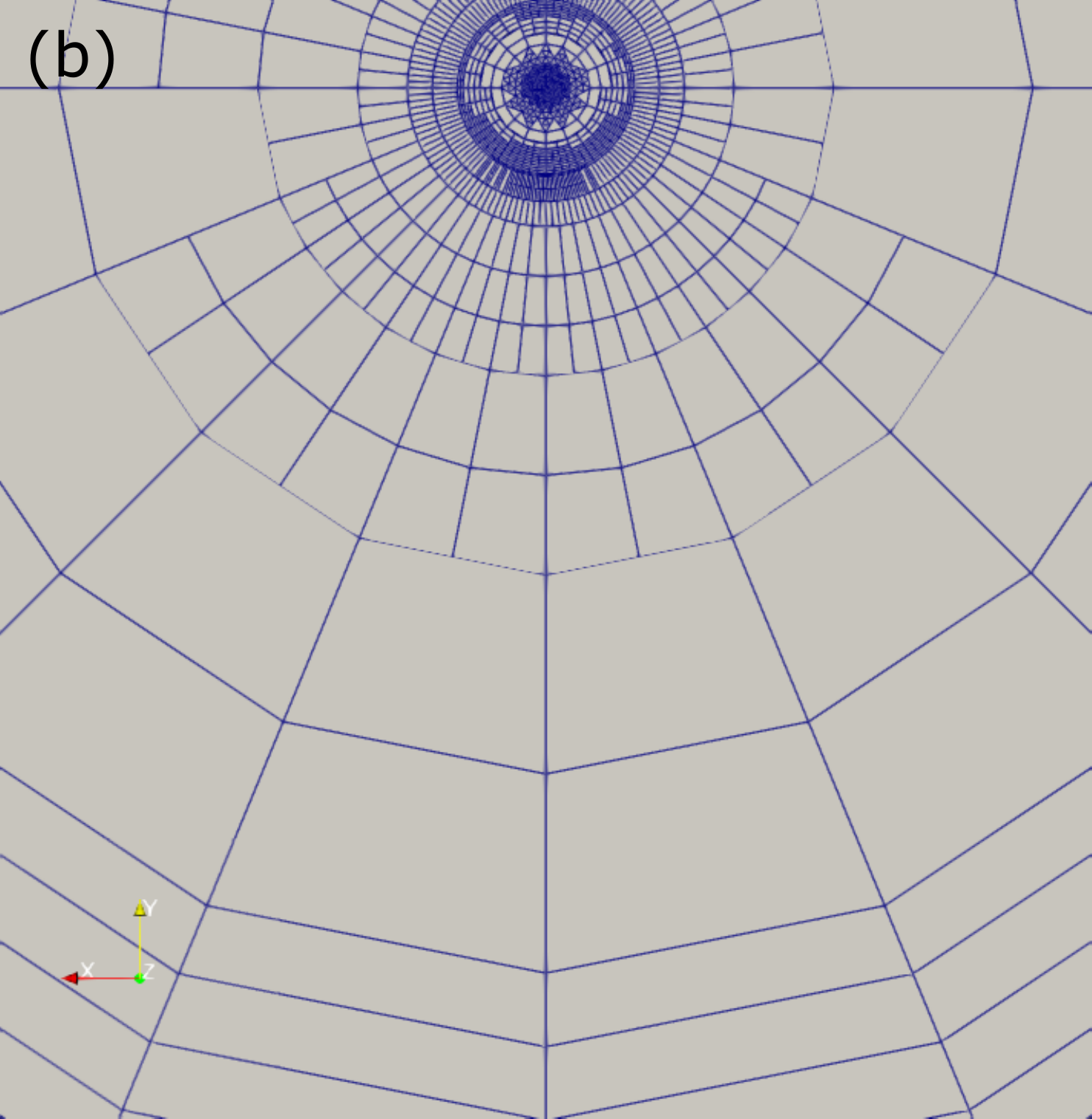}
    \end{minipage}
    \begin{minipage}{0.4865\textwidth}
    \includegraphics[width=\linewidth]{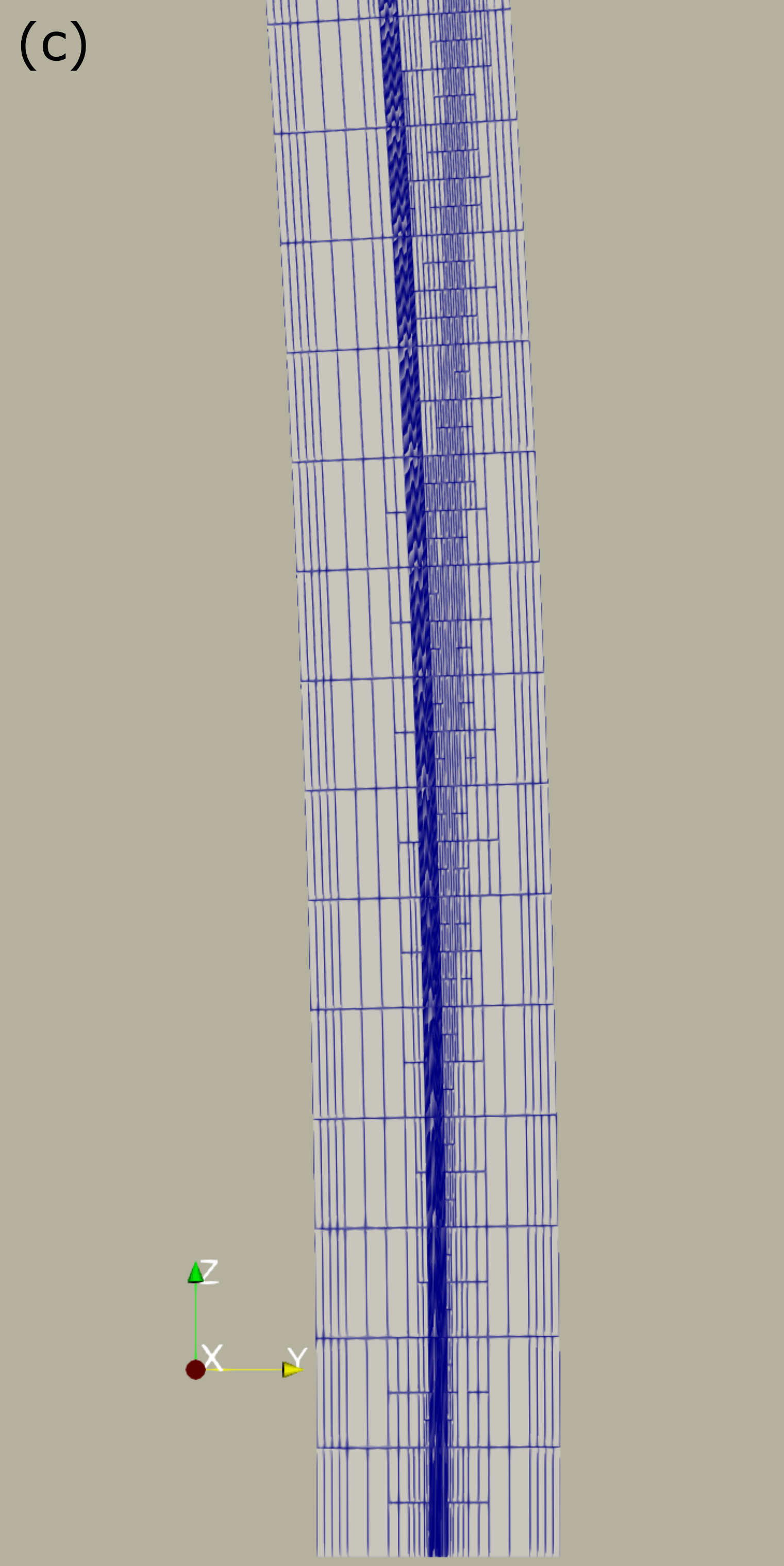}
    \end{minipage}
    \caption{Mesh plots of the final adapted mesh in the 3D bent fiber simulation (numerical experiment 3) with $\rbend=1300$: 
    (a)~3D view of the mesh; (b) cross-sectional slice at the fiber input facet ($\theta=0$); and (c)~longitudinal slice in the plane $x=0$. Note that the mesh plots were generated in ParaView using linear geometry outputs; the finite element computations use exact geometry elements.}
    \label{fig:ex3_mesh}
\end{figure}

\begin{figure}[htbp]
    \centering
    \includegraphics[width=0.495\linewidth]{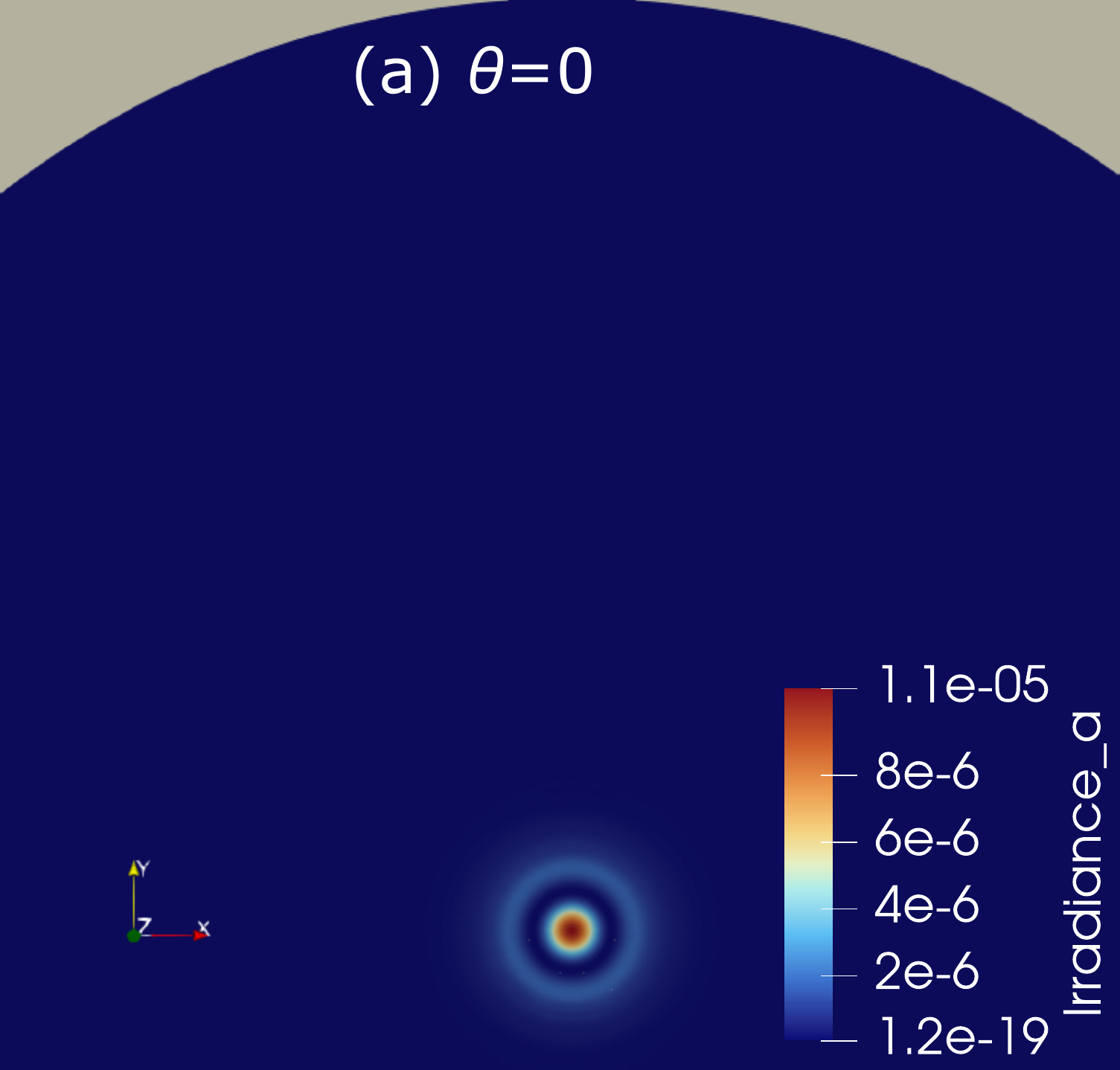}
    \includegraphics[width=0.495\linewidth]{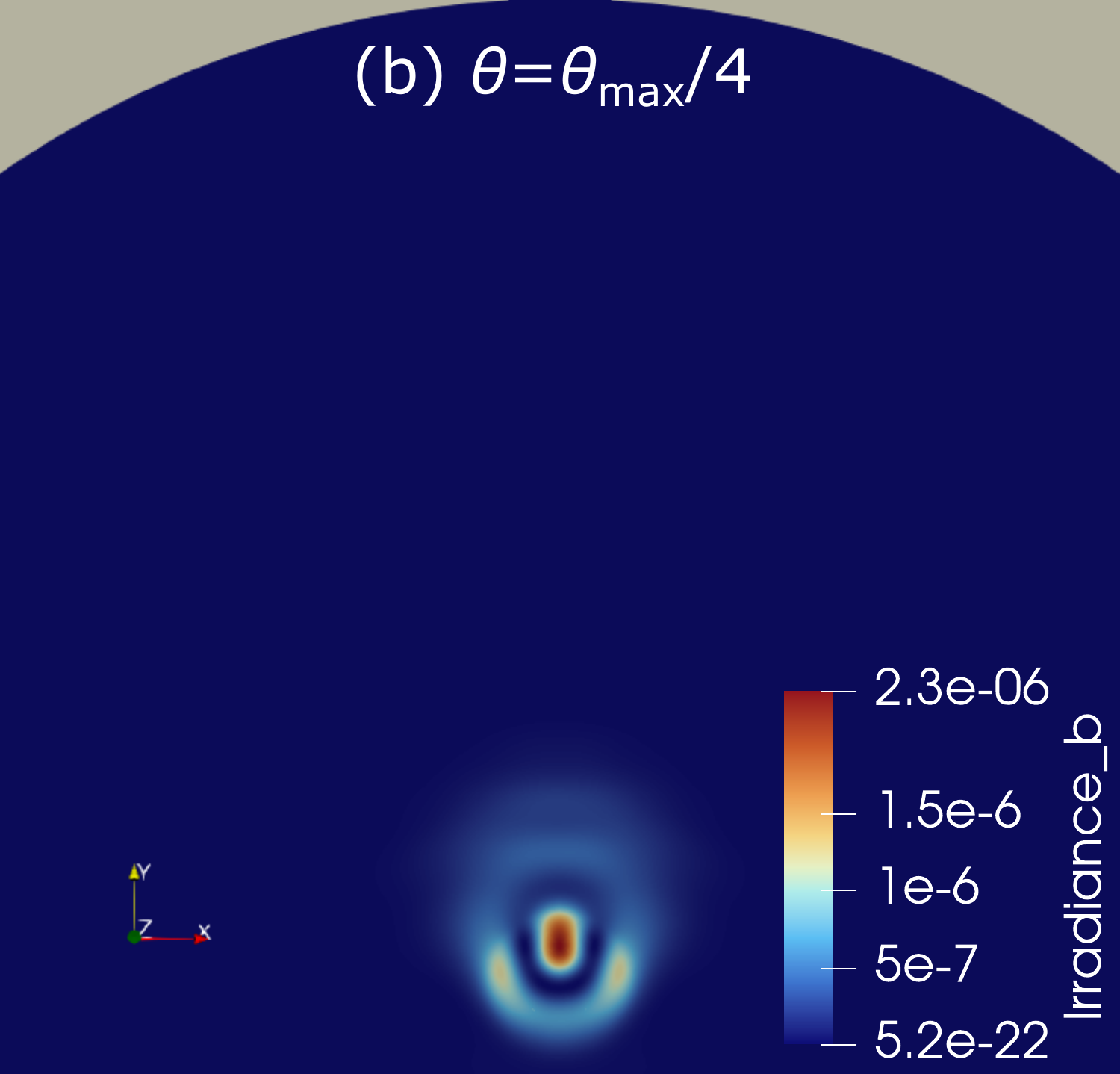}
    \includegraphics[width=0.495\linewidth]{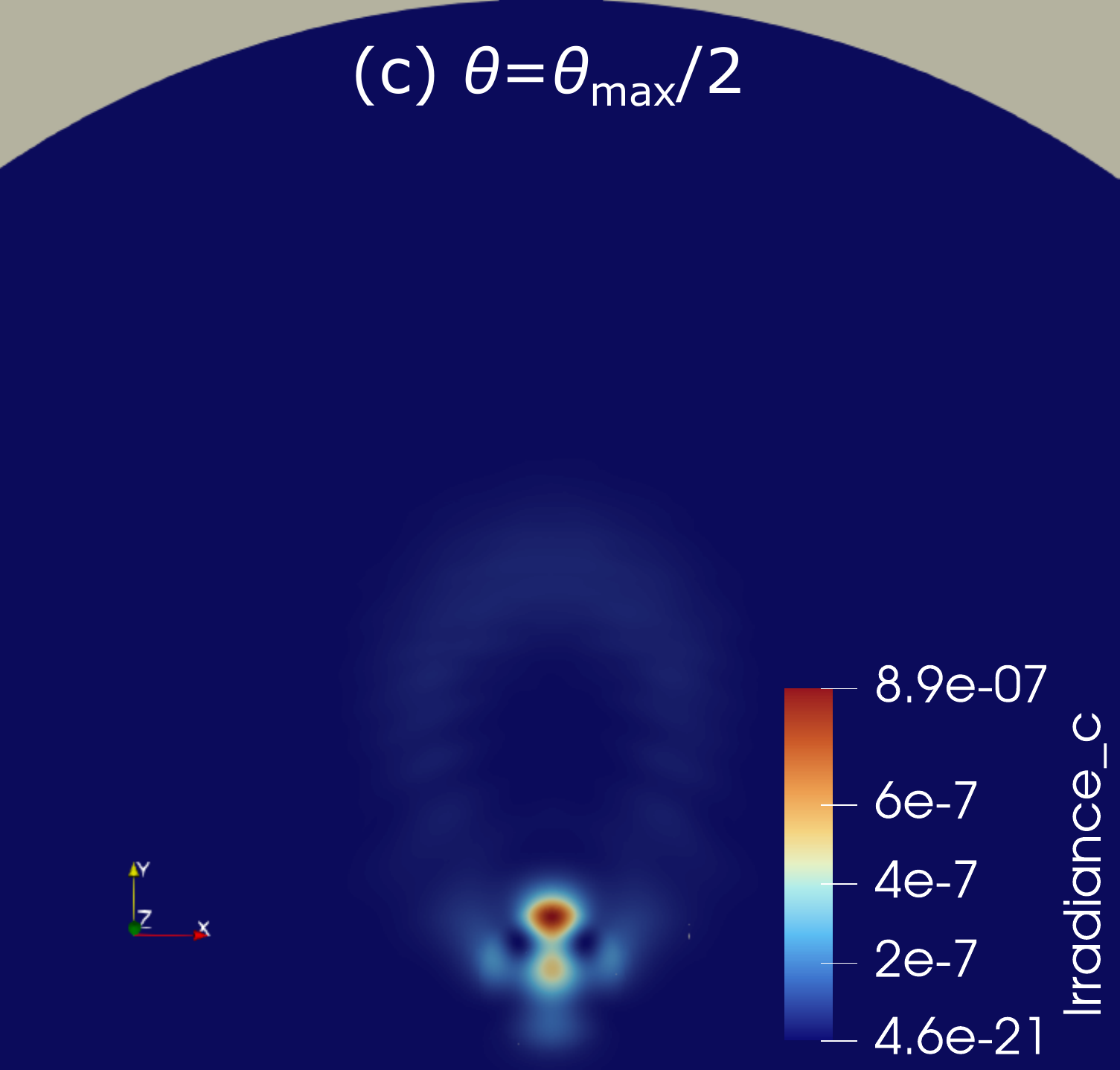}
    \includegraphics[width=0.495\linewidth]{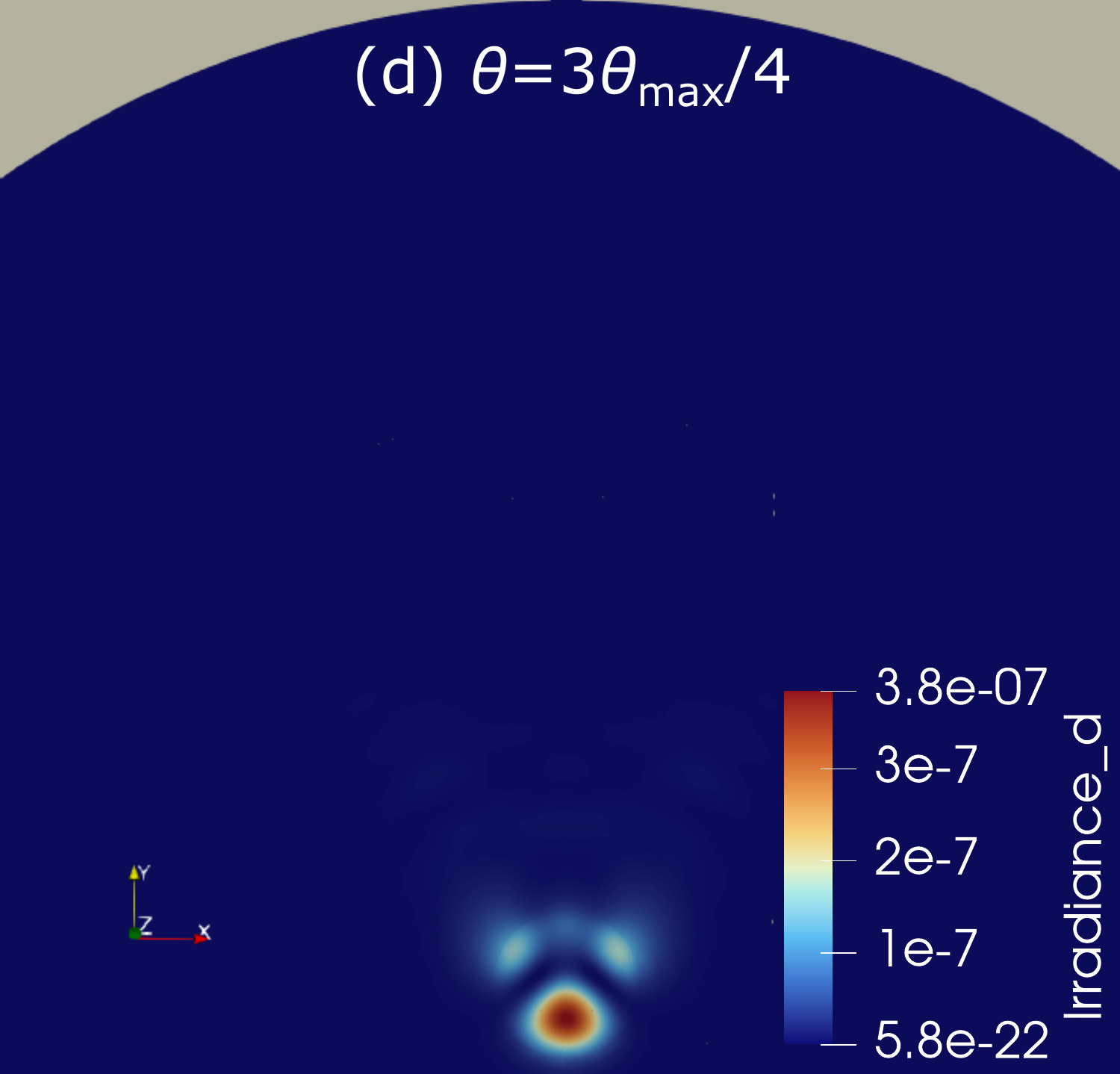}
    \caption{Results of the 3D bent fiber simulation (numerical experiment 3) with $\rbend=1300$. 
     Visualization of the irradiance field in the fiber core and cladding region at four different cross sections: (a)~$\theta=0$; (b)~$\theta=\theta_{\max}/4$; (c)~$\theta=\theta_{\max}/2$; and (d)~$\theta=3\theta_{\max}/4$. The input at the fiber input facet ($\theta=0$) is the LP$_{02}$ mode of the straight waveguide. Note that each picture has its own distinct color scale. As the light propagates down the bent fiber, the energy initially confined to the fiber core and inner cladding near the fiber core ($\theta=0$) leaves the core region and propagates toward the outer cladding where it is absorbed by the PML boundary.}
    \label{fig:ex3_slices}
\end{figure}

As a sample of the three-dimensional mesh and the resulting optical field, we present two sets of pictures, both corresponding to the case characterized by $\rbend=1300$.
The first set of images is grouped in Figure~\ref{fig:ex3_mesh}, where the final adaptively refined mesh can be appreciated from different views, including a longitudinal cut view. The second set of pictures shows cross-sectional \emph{slices} of the computational domain at four different values, $\theta \in\{ 0, 0.25, 0.5, 0.75 \} \theta_{\max}$. The visualized scalar field is the irradiance, post-processed from the numerical field solutions through \eqref{eq:te-irradiance1}. Notably, the optical field transverse profile varies strongly as it progresses along the bent fiber, from the core-guided (input) LP$_{02}$ mode (at $\theta = 0$) to patterns that progressively spread out of the fiber core region. We observe that not only \revS{does} the irradiance profile \revS{vary strongly} along the bent fiber, but its magnitude \revS{also changes substantially}, as illustrated by the successively changing color scales of the colorbars in the plots of \revS{Figure~\ref{fig:ex3_slices}}, with the maximum irradiance decaying nearly four orders of magnitude from $\theta = 0$ to $\theta = 0.75 \theta_{\max}$.

\begin{figure}[htb]
    \centering
    \includegraphics[width=0.5\textwidth]{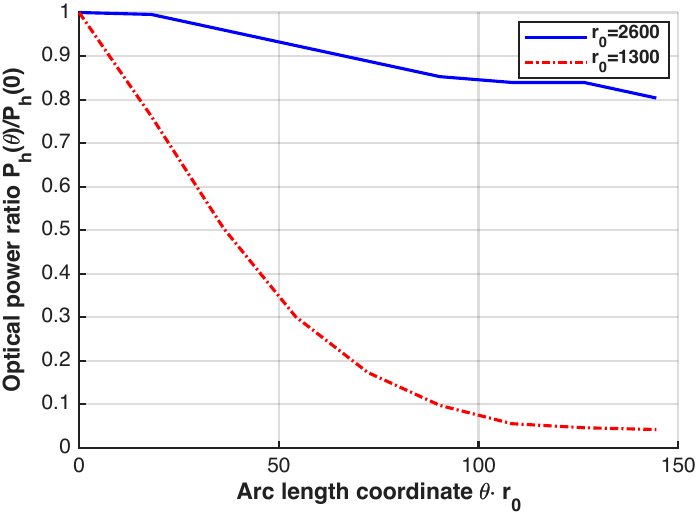}
    \caption{Results of the 3D bent fiber simulation (numerical experiment 3) with two different bend radii, $\rbend \in \{ 1300, 2600 \}$. The input at the fiber input facet ($\theta=0$) is the LP$_{02}$ mode of the straight waveguide. The plot shows the decay of the optical power, $\mcP_h(\theta)$, carried by this mode along the bent fiber as a function of the arc length coordinate. For a tighter bend radius, the power decays much more strongly, as expected.}
    \label{fig:ex3-power-results}
\end{figure}

We conclude this third numerical experiment by showing the relative value of the optical power (i.e., the ratio $\mcP_h(\theta)/\mcP_h(0)$, cf. \eqref{eq:discrete-optical-power}) as a function of the arc length coordinate, $\theta\cdot \rbend$, and compare the behavior for each of the bend radii
(see Figure~\ref{fig:ex3-power-results}). We note that, as expected, the optical power loss is significantly greater in the fiber configurations that use a smaller bend radius.



%% file: 5_conclusions.tex
\section{Conclusions}
\label{sec:conclusions}

In the presented work, we have extended the time-harmonic Maxwell envelope formulation to the bent waveguide/fiber setting. We constructed specialized PMLs to absorb the bend-induced radiation and presented a DPG finite element discretization of the ultraweak variational formulation corresponding to this model. 
Two numerical experiments, based on two-dimensional bent slab waveguides, provided evidence for error and residual convergence (\revisedA{Subsection}~\ref{subsec:experiment1}). The numerical tests also demonstrated the accurate \revS{computation} of physical quantities of interest, such as the bend loss exponent (cf.\ \eqref{eq:num-loss-factor}), \revS{which matches} the imaginary part of the eigenvalue of the corresponding Bessel differential equation, as discussed in \revisedA{Subsection}~\ref{subsec:experiment2}. 
The final numerical experiment in this paper \revS{considered} a fully three-dimensional setup, in which we analyzed the propagation of straight optical fiber modes as they enter a bent domain. 
The numerical results illustrated the distortion of the optical field profile due to geometric bending effects along the fiber, and the corresponding power loss of the previously guided mode for two different bend radii.

Our ongoing and future efforts are focused mainly on three different aspects.
On the computational side, we aim to simulate the onset of TMI and study its dependence on the curvature. 
One major challenge for this work is the deteriorating solver efficiency for larger-scale simulations of this high-frequency nonlinear wave problem. 
On the theoretical side, we aim to extend our 
well-posedness and stability analysis for circular waveguides with \revS{an} impedance boundary condition \cite{demkowicz2026length} to the case of a domain truncated \revS{by a PML that replaces} the radiation condition. 
Finally, we are working toward extending our current bending model by including elasto-optical effects\revS{, which} can be incorporated in the Maxwell model through a perturbation of the fiber refractive index that depends on the local bending strain~\cite{schermer2007improved}.